\documentclass[journal]{IEEEtran}
\usepackage{graphicx}
\usepackage{amsmath}
\ifCLASSINFOpdf
\else
\fi
\begin{document}
%
\title{Incident-Angle Dependence of Electromagnetic Wave Transmission through a Nano-hole in a Thin Plasmonic Semiconductor Layer}
%
%
%
\author{D\'{e}sir\'{e}~Miessein,
        Norman~J.~Morgenstern~Horing,
        Harry~Lenzing~and~Godfrey Gumbs
\thanks{D\'{e}sir\'{e}~Miessein is with the Department of Physics and Engineering Physics,
Stevens Institute of Technology, Hoboken, NJ 07030, USA (e-mail: dmiessei@stevens.edu)
(present address: Department of Physics and Engineering Physics, Fordham University, Bronx, NY 10458, USA )}
\thanks{Norman~J.~Morgenstern~Horing is with the Department of Physics and Engineering Physics,
Stevens Institute of Technology, Hoboken, NJ 07030, USA
(phone:201-216-5651; fax:201-216-5638; e-mail: nhoring@stevens.edu)}
\thanks{Harry Lenzing is with the Department of Physics and Engineering Physics,
Stevens Institute of Technology, Hoboken, NJ 07030, USA (e-mail: harry07757@comcast.net)}
\thanks{Godfrey Gumbs is with the Department of Physics and Astronomy,
Hunter College of the City University of New York, New York, NY 10065, USA (e-mail: ggumbs@hunter.cuny.edu)}
}

\date{\today}
%
%
%
\maketitle
%
\begin{abstract}
This work is focussed on the role of the angle of incidence of an incoming electromagnetic wave in its transmission
through a subwavelength nano-hole in a thin semiconductor plasmonic layer.  Fully detailed calculations and results are exhibited for $ p$-
and $s$-polarizations of the incident wave for a variety of incident angles in the near, middle and far zones of the transmitted radiation.
Our dyadic Green's function formulation includes both (1) the electromagnetic field transmitted directly through the $ 2D $ plasmonic layer
superposed with (2) the radiation emanating from the nano-hole.  Interference fringes due to this superposition are explicitly exhibited.
Based on an integral equation formulation, this dyadic Green's function approach does not involve any appeal to metallic boundary conditions.
It does incorporate the role of the $ 2D $ plasmon of the semiconductor layer, which is smeared due to its lateral wave number dependence.

We find that the interference fringes, which are clustered near the nano-hole, flatten to a uniform level of transmission directly
through the sheet alone at large distances from the nano-hole.  Furthermore, as the incident angle increases, the axis of the relatively
large central transmission maximum through the nano-hole follows it, accompanied by a spatial compression of interference fringe maxima forward of
the large central transmission maximum, and a spatial thinning of the fringe maxima behind it.  For $p$-polarization, the transmission results show
a strong increase as the incident angle $\theta_0 $ increases, mainly in the dominant $ E_z$ component (notwithstanding a concomitant decrease
of the $ E_x $ component as $\theta_0$ increases).  We also find that in the case of $s$-polarization of the incident electromagnetic wave,
the transmission decreases as $\theta_0$ increases.  These results, for both $p$- and $s$-polarizations, are consistent  with earlier  results
for perfect metal boundary conditions, which are \emph{not invoked here} as we have treated the problem of a nano-hole in a \emph{semiconductor} layer and
have determined its electromagnetic transmission including the role of its two dimensional plasma.
\end{abstract}
%
%
%
%
\IEEEpeerreviewmaketitle
\section{Introduction}

%
%
%
%
\IEEEPARstart{I}{n} a recent study of electromagnetic wave transmission through a nano-hole in a thin plasmonic semiconductor layer,
we examined the transmission in detail for normal incidence of the wave train [1-5].  Our dyadic Green's function formulation elliminated the need
for separate treatment of the boundary conditions used in earlier works [6].  Other important works [7-9] using dyadic Green's functions and variational
principles assumed the layer to be a perfect metallic conductor, whereas our present interest considers the layer to be a thin plasmonic semiconductor.
Similar remarks apply to additional interesting studies [10-19] that restrict consideration to a layer that is a perfect metal, which restriction is eliminated
in the present work that also treats arbitrary angles of incidence of the incoming electromagnetic wave.

Our earlier work [1] involved the analytic determination
of the dyadic electromagnetic Green's function for the system in closed form, facilitating relatively simple calculation
of the transmitted radiation through both the hole as well as through the layer itself for normal incidence. Employment of the Green's function
in an integral equation formulation automatically embedded the role of the electromagnetic boundary conditions and the 2D plasmon
in the layer (which is "smeared" due to its lateral wavenumber dependence) obviating the need for separate treatment.
In the present paper, we employ the same method to analyze electromagnetic transmission through such a system for \emph{non}-normal
incidence of the electromagnetic wave train.  In Section 2, we review the appropriate dyadic Green's function for the perforated
plasmonic screen with a subwavelength aperture and its application to electromagnetic transmission in the construction of the
system's inverse dielectric tensor, with emphasis on non-normal incidence. The calculated results for various angles of incidence
are exhibited in the figures of Section 3, and a summary is presented in Section 4.
%
%
%
%
%
\section{ Dyadic Green's Function and  Inverse Dielectric Tensor }
\numberwithin{figure}{section}
\begin{figure}[h]
\centering
\includegraphics[width=7cm,height=8cm]{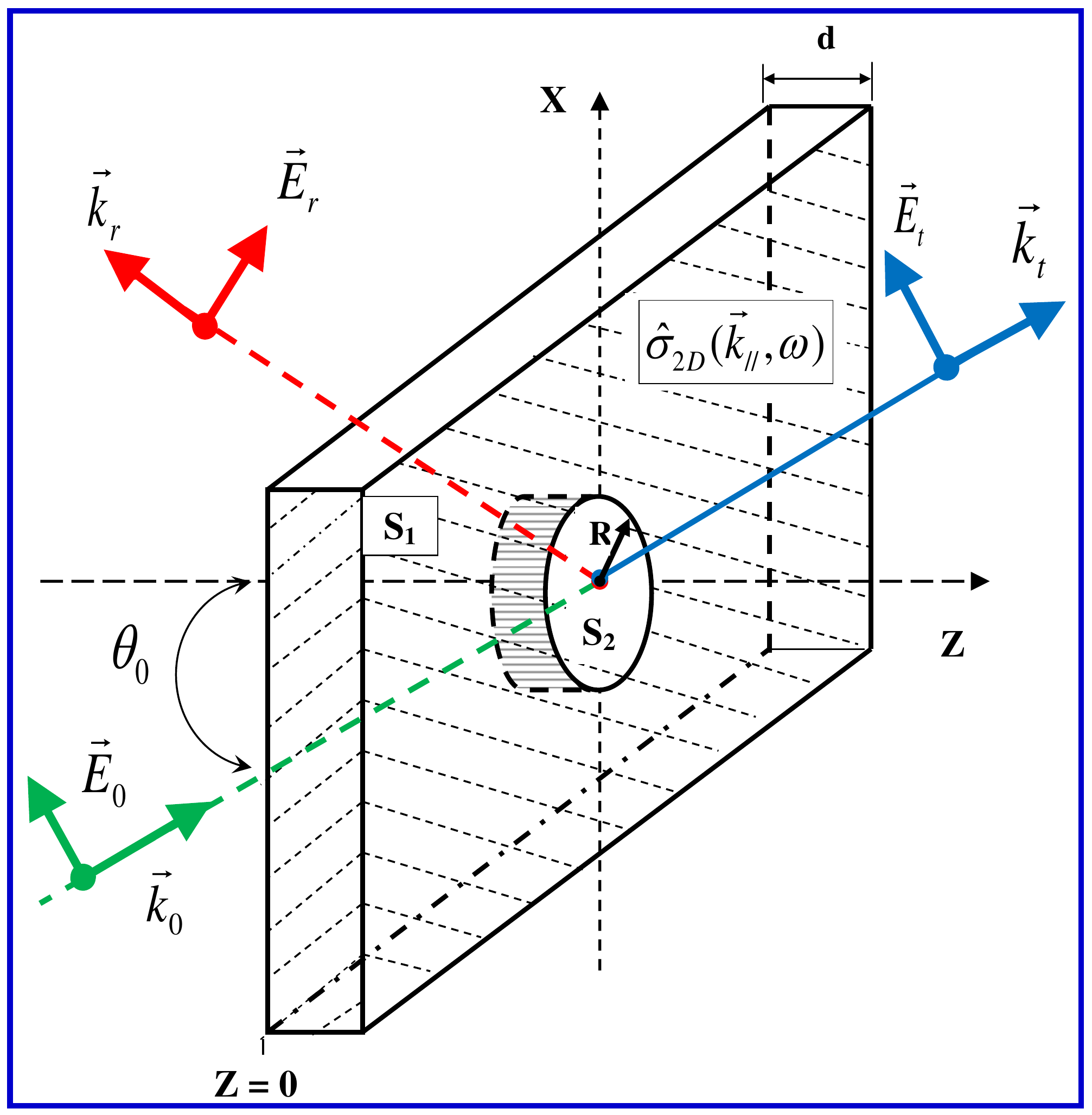}
\caption{Two dimensional plasmonic layer (thickness $ d $, embedded
at $ z=0 $ in a three dimensional bulk medium) with a nano-hole of radius
$ R $, area $ A $ at the origin of the $(x-y)$ plane., shown with incident, reflected and
transmitted wave vectors ($ \vec{k}_{0}, \vec{k}_{r}, \vec{k}_{t} $) for waves
($ \vec{E}_{0}(x,y,z;t), \vec{E}_{r}(x,y,z;t), \vec{E}_{t}(x,y,z;t)$).  The angle of incidence is $ \theta_{0}$
 in the $x-z$ plane($ k_{0y}\equiv 0 $).}
\label{FIG1411}
\qquad
\end{figure}
The dyadic Green's function for a thin perforated plasmonic screen with a nano-hole was determined
as $ \left( \beta = \gamma {A}=\frac{4\,i\pi\omega}{c^2}\sigma^{(2D)}_{fs}(\omega)\,{A} \right)$:
\begin{eqnarray}\label{A2.1}
\widehat{G}(\vec{r}_{\parallel},0;z,0;\omega)&=&
 {\widehat{G}_{fs}(\vec{r}_{\parallel},0;z,0;\omega)}
 \nonumber\\
 &\times &
\left[ {\widehat{I}\,+\,\beta \widehat{G}_{fs}(0,0;0,0;\omega)}\right]^{-1}.
\end{eqnarray}
\normalsize
Here, $\widehat{G}_{fs}$ is the dyadic Green's function for the thin plasmonic layer in the absence of the
subwavelength aperture, and it is given by
\tiny
\begin{eqnarray}\label{A2.2}
  {\widehat{{G}}_{fs}}({k}_{x},{k}_{y}=0;z,0;\omega) =
  \nonumber\\
  \begin{bmatrix}
  {G}_{fs}^{xx}({k}_{x},{k}_{y}=0;z,0;\omega) & 0 & {G}_{fs}^{xz}({k}_{x},{k}_{y}=0;z,0;\omega) \\
  0 & {G}_{fs}^{yy}({k}_{x},{k}_{y}=0;z,0;\omega) & 0 \\
  {G}_{fs}^{zx}({k}_{x},{k}_{y}=0;z,0;\omega) & 0 & {G}_{fs}^{zz}({k}_{x},{k}_{y}=0;z,0;\omega)
 \end{bmatrix}
 \nonumber\\
\end{eqnarray}
\normalsize
where $\vec{k}_{\parallel}$ is wavevector parallel to the plasmonic screen and the matrix elements of
$ \widehat{G}_{fs}$ are given by
\begin{equation}\label{A2.3}
{G}_{fs}^{xx}({k}_{x},{k}_{y}=0;z,0;\omega)=\,-\,\frac{e^{i\,k_{z}\mid z\mid }}{2\,i\,k_{z}}\left[\frac{1}{\overline{D}_{1}}\left\{\left(1-\frac{k_{x}^{2}}{q_{\omega}^{2}}\right) \right\}\right],
\end{equation}
\begin{equation}\label{A2.4}
{G}_{fs}^{yy}({k}_{x},{k}_{y}=0;z,0;\omega)=\,-\,\frac{e^{i\,k_{z}\mid z\mid }}{2\,i\,k_{z}}\left[\frac{1}{\overline{D}_{3}}\right],
\end{equation}
\small
\begin{eqnarray}\label{A2.5}
{G}_{fs}^{zz}({k}_{x},{k}_{y}=0;z,0;\omega)&=&-\frac{e^{i\,k_{z}\mid z\mid }}{2\,i\,k_{z}}
\nonumber\\
&\times &
\left[\frac{1}{\overline{D}_{2}}\left\{\left(1-\frac{k_{z}^{2}- 2 i k_{z} \delta (z)}{q_{\omega}^{2}}\right)\right\}\right],
\nonumber\\
\end{eqnarray}
\normalsize
\begin{equation}\label{A2.6}
{G}_{fs}^{xy}({k}_{x},{k}_{y}=0;z,0;\omega)=0,
\end{equation}
\small
\begin{equation}\label{A2.7}
{G}_{fs}^{xz}({k}_{x},{k}_{y}=0;z,0;\omega)=\,+\,\frac{e^{i\,k_{z}\mid z\mid }}{2\,i\,k_{z}}\left[\frac{1}{\overline{D}_{2}}\left(\frac{k_{x}\,k_{z}sgn(z)}{q_{\omega}^{2}}\right)\right],
\end{equation}
\normalsize
\begin{equation}\label{A2.8}
{G}_{fs}^{yz}({k}_{x},{k}_{y}=0;z,0;\omega)=0,
\end{equation}
\begin{equation}\label{A2.9}
{G}_{fs}^{yx}({k}_{x},{k}_{y}=0;z,0;\omega) = 0,
\end{equation}
\small
\begin{equation}\label{A2.10}
{G}_{fs}^{zx}({k}_{x},{k}_{y}=0;z,0;\omega)=\,+\,\frac{e^{i\,k_{z}\mid z\mid }}{2\,i\,k_{z}}\left[\frac{1}{\overline{D}_{1}}\left(\frac{k_{z}\,k_{x}sgn(z)}{q_{\omega}^{2}}\right)\right],
\end{equation}
\normalsize
\begin{equation}\label{A2.11}
{G}_{fs}^{zy}({k}_{x},{k}_{y}=0;z,0;\omega)=0,
\end{equation}
and
\begin{small}
\begin{equation*}
\bigg( \gamma\,=\,\frac{4\pi\,i\,\omega}{c^{2}}\,{\sigma}_{fs}^{(2D)}({\vec{k}}_{\parallel};\omega); \: {\sigma}_{fs}^{(2D)}({\vec{k}}_{\parallel};\omega)=\frac{i\,\omega}{4\pi}\left[ \varepsilon_{b}^{(3D)}-\varepsilon(\omega)\right]d \bigg)
\end{equation*}
\end{small}
\begin{equation}\label{A2.12}
   \overline{D}_{1}=\left[1+ \left(\frac{\gamma}{2\,i\,k_{z}}\right) \left( 1- \frac{k_{x}^{2}}{q_{\omega}^{2}}\right) \right],
\end{equation}
\begin{equation}\label{A2.13}
   \overline{D}_{2}=\left[1+ \left(\frac{\gamma}{2\,i\,k_{z}}\right) \left( 1- \frac{a_{0}^{2}}{q_{\omega}^{2}}\right) \right],
\end{equation}
\begin{equation}\label{A2.14}
   \overline{D}_{3}=\left[1+ \left(\frac{\gamma}{2\,i\,k_{z}}\right)\right]
\end{equation}
with $ a_{0}^{2}=k_{z}^{2}\,-\,\frac{2\,i\,k_{z}}{d} $.

While the dyadic Green's function above determines the electromagnetic response to a current source, it does not, by itself,
describe the response to an incident electromagnetic wave field.  For that, it is necessary to construct the inverse
dielectric tensor of the system  which provides the relation between the actual field $ \vec{E}(\vec{r};t)$ and the impressed (incident)
field $ \vec{E}_{0}(\vec{r};t)$.  Considering that $ \widehat{G}(\vec{r}_{\parallel},\vec{r}^{\,\prime}_{\parallel};z,z^{\,\prime};\omega)$ described above already
incorporates the role of induced current, it provides the system's response to an externally impressed current $ \vec{J}_{ext}$
\textit{alone} as
\begin{eqnarray}\label{A2.15}
\vec{E}(\vec{r}_{\parallel},z;\omega)&=&\frac{4\pi\,i\,\omega}{c^{2}}
\int d^{2}\vec{r}^{\,\prime}_{\parallel}
\nonumber\\
&\times &
\int d\vec{z}^{\,\prime}\,{\widehat{G}}(\vec{r}_{\parallel},z;\vec{r}^{\,\prime}_{\parallel},z^{\prime};\omega)\,\vec{J}_{ext}(\vec{r}^{\,\prime}_{\parallel},z^{\,\prime};\omega)
\nonumber\\
\end{eqnarray}
or
\begin{equation}\label{A2.16}
\vec{E}\,=\,\frac{4\pi\,i\,\omega}{c^{2}}\,{\widehat{G}}\,\vec{J}_{ext},
\end{equation}
in matrix notation.
Bearing in mind that ($ \widehat{\sigma}^{(2D)}_{fs}$, $\widehat{\sigma}^{(2D)}_{hole} $ are the conductivities of the full 2D screen and the excluded hole, respectively)
\begin{eqnarray}\label{A2.17}
\widehat{G}^{-1}= \widehat{G}^{-1}_{3D}
& - & \frac{4\pi\,i\omega}{c^{2}}\,\widehat{\Sigma}_{2D}
\end{eqnarray}
where
\begin{equation}\label{A2.18}
\widehat{\Sigma}_{2D}=
{\widehat{\sigma}_{fs}}^{(2D)}\,-\,
{\widehat{\sigma}_{hole}}^{(2D)}
\end{equation}
and that the incident field $ \vec{E}_{0}$ is related to its distant current source $ \vec{J}_{ext}$ by
\begin{equation}\label{A2.19}
\vec{E}_{0}\,=\,\frac{4\pi\,i\,\omega}{c^{2}}\,{\widehat{G}_{3D}}\,\vec{J}_{ext}
\end{equation}
or
\begin{equation}\label{A2.20}
\vec{J}_{ext}\,=\,\left[\frac{4\pi\,i\,\omega}{c^{2}}\,{\widehat{G}_{3D}}\right]^{-1}\vec{E}_{0},
\end{equation}
we have
\begin{eqnarray}\label{A2.21}
\vec{E}\,&=&\,\widehat{G}\,\widehat{G}^{-1}_{3D}\vec{E}_{0}
\nonumber\\
&=&\,
\widehat{G}\left[\widehat{G}^{-1}\,+\,\frac{4\pi\,i\,\omega}{c^{2}}\,{\widehat{\Sigma}_{2D}}\right]\vec{E}_{0}
\nonumber\\
&=&\,
\left[\widehat{I}\,+\,\frac{4\pi\,i\,\omega}{c^{2}}\widehat{G}\,{\widehat{\Sigma}_{2D}}\right]\vec{E}_{0}.
\end{eqnarray}
Obviously, this introduces the inverse dielectric dyadic/tensor $\widehat{K}$ as
\begin{eqnarray}\label{A2.22}
\widehat{K}\,=\,
\left[\widehat{I}\,+\,\frac{4\pi\,i\,\omega}{c^{2}}\widehat{G}\,{\widehat{\Sigma}_{2D}}\right],
\end{eqnarray}
in complete analogy to the results\cite{Levine_1950,Levine_1948} of references \cite{Dez_2015} and \cite{Dez_thesis}.

Thus, we have the actual $\vec{E}$-field as
\begin{eqnarray}\label{A2.23}
\vec{E}(\vec{r}_{\parallel};z;\omega) =
\vec{E}_{0}(\vec{r}_{\parallel};z;\omega)\,+\,\vec{E}_{1}(\vec{r}_{\parallel},z;\omega)\,+\,\vec{E}_{2}(\vec{r}_{\parallel},z;\omega),
\nonumber\\
\end{eqnarray}
where the electric field contributions $\vec{E}_{1}$ and $\vec{E}_{2}$ are defined by
\begin{eqnarray}\label{A2.24}
\vec{E}_{1}(\vec{r}_{\parallel},z;\omega)& =&
\frac{4\pi\,i\,\omega}{c^{2}}\int{d^{2}\vec{r}_{\parallel}^{'}}
\nonumber\\
&\times&
\int{d^{2}\vec{r}_{\parallel}^{''}} \int{dz^{'}} \int{dz^{''}}
\widehat{G}(\vec{r}_{\parallel},\vec{r}_{\parallel}^{'};z,z^{'};\omega)
\nonumber\\
&\times&
{\widehat{\sigma}_{fs}}^{(2D)}
(\vec{r}_{\parallel}^{'},\vec{r}_{\parallel}^{''};z^{'},z^{''};\omega)\,
\vec{E}_{0}(\vec{r}_{\parallel}^{''};z^{''};\omega)
\nonumber\\
\end{eqnarray}
and
\begin{eqnarray}\label{A2.25}
\vec{E}_{2}(\vec{r}_{\parallel},z;\omega) &=&
\frac{4\pi\,i\,\omega}{c^{2}}\int{d^{2}\vec{r}_{\parallel}^{'}}
\nonumber\\
&\times&
\int{d^{2}\vec{r}_{\parallel}^{''}} \int{dz^{'}} \int{dz^{''}}
\widehat{G}(\vec{r}_{\parallel},\vec{r}_{\parallel}^{'};z,z^{'};\omega)
\nonumber\\
&\times&
{\widehat{\sigma}_{hole}}^{(2D)}
(\vec{r}_{\parallel}^{'},\vec{r}_{\parallel}^{''};z^{'},z^{''};\omega)
\vec{E}_{0}(\vec{r}_{\parallel}^{''};z^{''};\omega).
\nonumber\\
\end{eqnarray}
Because the conductivities $ \widehat{\sigma}^{2D}_{fs} $ and $ \widehat{\sigma}^{2D}_{hole}$ are confined to the 2D screen
at $ z=0 $ (thickness $d$) they may be written in lateral and $ z $ - representation as
\begin{equation}\label{A2.26}
   \widehat{\sigma}_{f_{s}}^{2D}(\vec{r}_{\parallel}^{'},\vec{r}_{\parallel}^{''};z^{'},z^{''};\omega)=
\widehat{I}\, {\sigma}_{fs}^{(2D)}(\omega)\delta^{(2D)}
(\vec{r}_{\parallel}^{'}-\vec{r}_{\parallel}^{''})\,\delta(z^{'})\,\delta(z^{''})
\end{equation}
and
\begin{eqnarray}\label{A2.27}
\widehat{\sigma}^{(2D)}_{hole}(\vec{r}_{\parallel},\vec{r}_{\parallel}^{'};z,z^{'};\omega)&\approx &
\widehat{I} {A} \sigma_{fs}^{(2D)}(\omega)
\nonumber\\
&\times&
\delta^{2D}(\vec{r}_{\parallel})\delta^{2D}(\vec{r}_{\parallel}-\vec{r}_{\parallel}^{'})\delta(z)\delta(z').
\nonumber\\
\end{eqnarray}
It should be noted that in the absence of the nano-hole,
\newline
$ A \longmapsto 0$, the field contribution
$ \vec{E}_{2}(\vec{r}_{\parallel},z;\omega)$ vanishes, leaving
\begin{eqnarray}\label{A2.28}
\vec{E}(\vec{r}_{\parallel};z;\omega) &=&
\vec{E}_{0}(\vec{r}_{\parallel};z;\omega)\,+\,\vec{E}_{1}(\vec{r}_{\parallel},z;\omega)
\nonumber\\
\,&=&\,\vec{E}_{fs}(\vec{r}_{\parallel},z;\omega),
\end{eqnarray}
where $ \vec{E} \longmapsto \vec{E}_{fs} $ jointly with $ \widehat{G} \longmapsto \widehat{G}_{fs}$
are the field and associated Green's function (respectively) for the transmission/reflection of the field
$ \vec{E}_{0}$ impingent on the full, unperforated 2D layer.

Finally, the resulting electric field $ \vec{E}$ can be written in position representation as
\begin{eqnarray}\label{A2.29}
\vec{E}(\vec{r}_{\parallel};z;t)&=&
\vec{E}_{0}(\vec{r}_{\parallel};z;t)
\nonumber\\
&+&
\gamma_{0}\,
\widehat{{G}}_{fs}(\vec{k}_{0_{\parallel}};z,0;\omega_{0})\,\vec{E}_{0}\,
e^{i\,\left[\,\vec{k}_{0_{\parallel}}\,\cdot\,\vec{r}_{\parallel}\,-\,\omega_{0}\,t\right]}\,
\nonumber\\
&-&
\,\beta_{0}\,\widehat{G}(\vec{r}_{\parallel},0;z,0;\omega_{0})
\nonumber\\
& \times &
\left[{\widehat{I}+\gamma_{0}
\widehat{\overline{\overline{G}}}_{fs}(\vec{k}_{0_{\parallel}};0,0;\omega_{0})}\right]
\,\vec{E}_{0}\,\,e^{-i\,\omega_{0}\,t}
\nonumber\\
\end{eqnarray}
where ($ \gamma_{0}=\,-\,\frac{d\,\omega_{P_{3D}}^{2}}{c^{2}}\,\,\,\, \textrm{and} \,\,\,\, \omega_{_{P3D}}= \sqrt{\frac{4\,\pi\,e^{2}\,\rho_{_3D}}{m^{\star}}}$
is the 3D bulk plasma frequency; $ \beta_{0} = \gamma_{0}\, {A} $; $ \Gamma_{0}=-\,\frac{i\,\gamma_{0}}{2}$; $ \vec{k}_{0}$ and $ \omega_{0}$ are the incident wavevector and frequency, respectively)
\begin{eqnarray}\label{A2.30}
\overline{\overline{G}}_{fs}^{\,xx}(\vec{k}_{0_{\parallel}};0,0;\omega_{0})=\, \frac{\,-\,\cos(\theta_{0})}{2\,i\,\left[q_{\omega_{0}}\,+\,\Gamma_{0}\,\cos(\theta_{0})\right]},
\end{eqnarray}
\begin{eqnarray}\label{A2.31}
\overline{\overline{G}}_{fs}^{\,yy}(\vec{k}_{0_{\parallel}};0,0;\omega_{0})=\,\frac{\,-\,1}{2\,i\,\left[\,\Gamma_{0}\,+\,q_{\omega_{0}}\,\cos(\theta_{0})\right]},
\end{eqnarray}
and

\begin{eqnarray}\label{A2.32}
\overline{\overline{G}}_{fs}^{\,zz}(\vec{k}_{0_{\parallel}};0,0;\omega_{0})=
\nonumber\\
\frac{\,-\,1}{2\,i}
\left\{\frac{q_{\omega_{0}}\,\sin^{2}(\theta_{0})+2\,i\,\delta(0)\cos(\theta_{0})}{\cos(\theta_{0})\,
\left[\,q_{\omega_{0}}^{2}\,+\,2\,i\,\delta(0)\Gamma_{0}\right]\,+\,\Gamma_{0}\,q_{\omega_{0}}\,\sin^{2}(\theta_{0})}\right\}.
\nonumber\\
\end{eqnarray}
\normalsize
(Note that $ \widehat{\overline{\overline{G}}}_{fs}(\vec{k}_{0_{\parallel}};0,0;\omega_{0})$ was shown to be diagonal
in reference \cite{Dez_2015}, see appendices)

\section{Incident Angle Dependence of Electromagnetic Wave Transmission Through a 2D Plasmonic Layer with a Nano-hole }
\begin{figure}[h]
\centering
(a)\\
 \includegraphics[width=9cm,height=7cm]{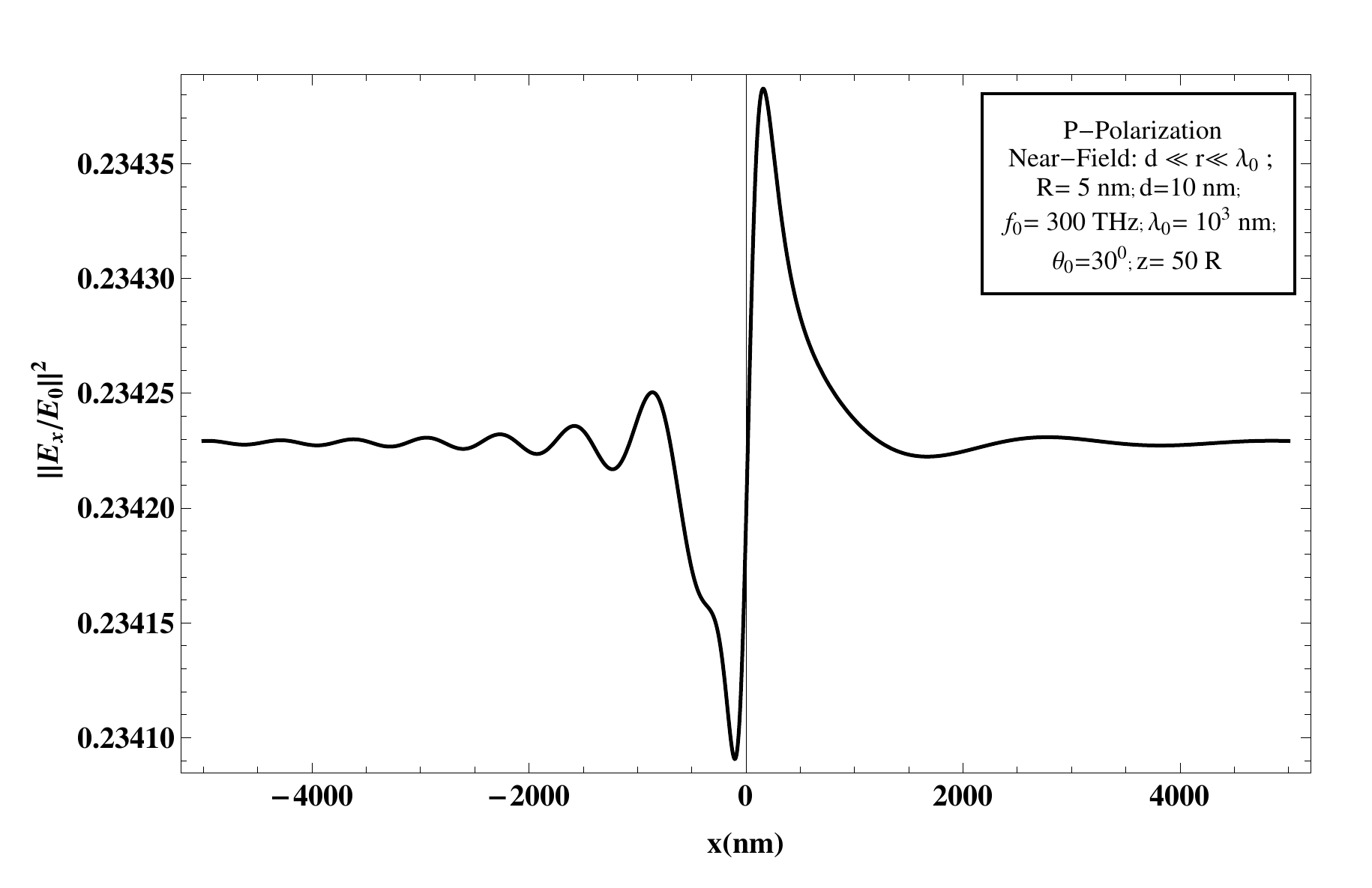}\\
(b)\\
 \includegraphics[width=9cm,height=7cm]{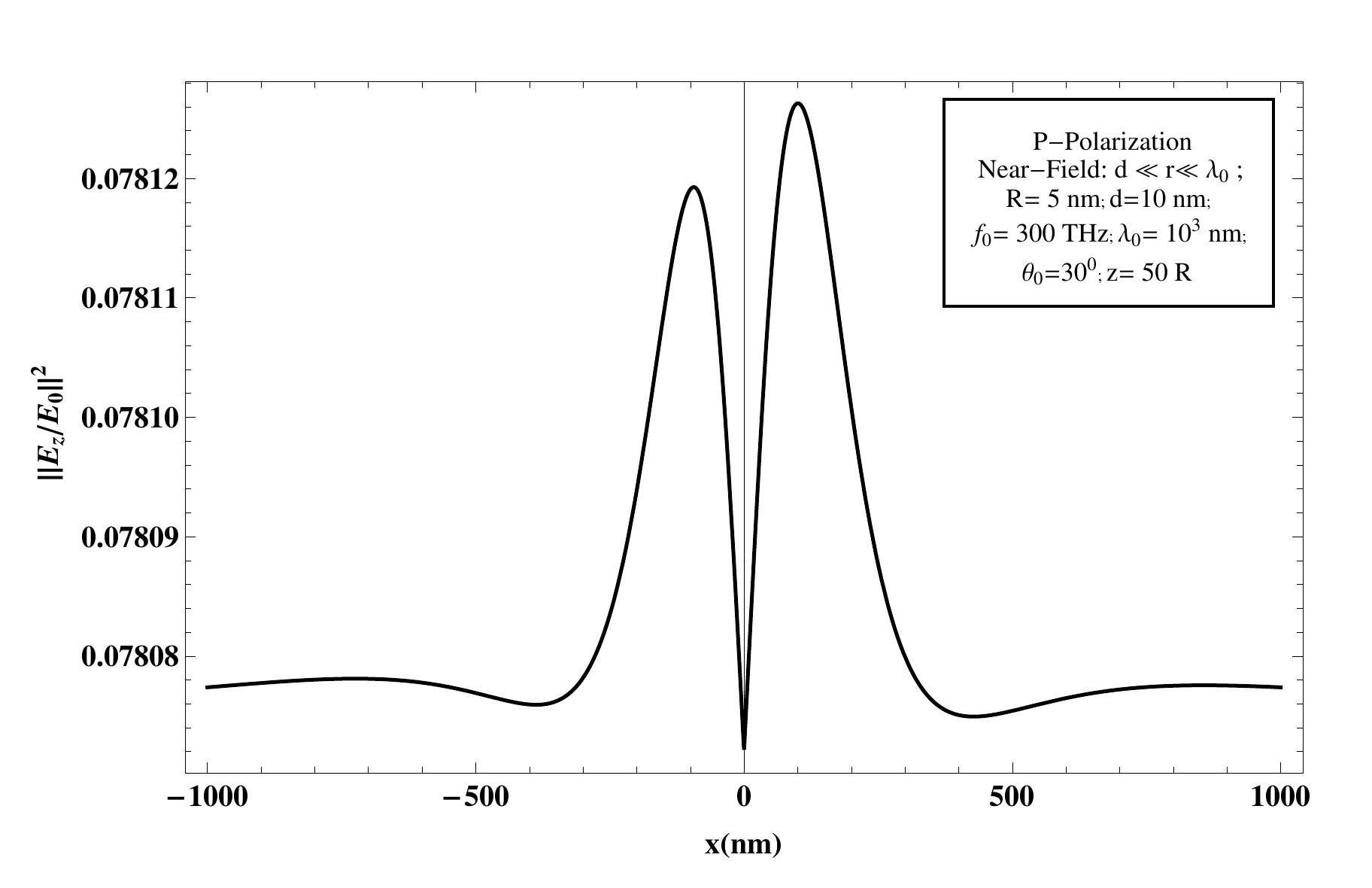}
\caption{ $p$-polarization-Near-Field, $ z = 50\,R $; \textbf{$ \theta_{0}=30^{\circ}$}:
\newline (a) $\mid {E}_{x}(x,y,z;t)/{E}_{0}\mid^{2}$  and (b) $\mid {E}_{z}(x,y,z;t)/{E}_{0}\mid^{2}$ produced by a perforated
2D plasmonic layer of GaAs as a function of lateral distance $ r_{_{\parallel}}= x\,(y=0)$ from the aperture.}
\label{FIGNFT30ExEz}
\qquad
\end{figure}

\begin{figure}[h]
\centering
(a)\\
 \includegraphics[width=9cm,height=7cm]{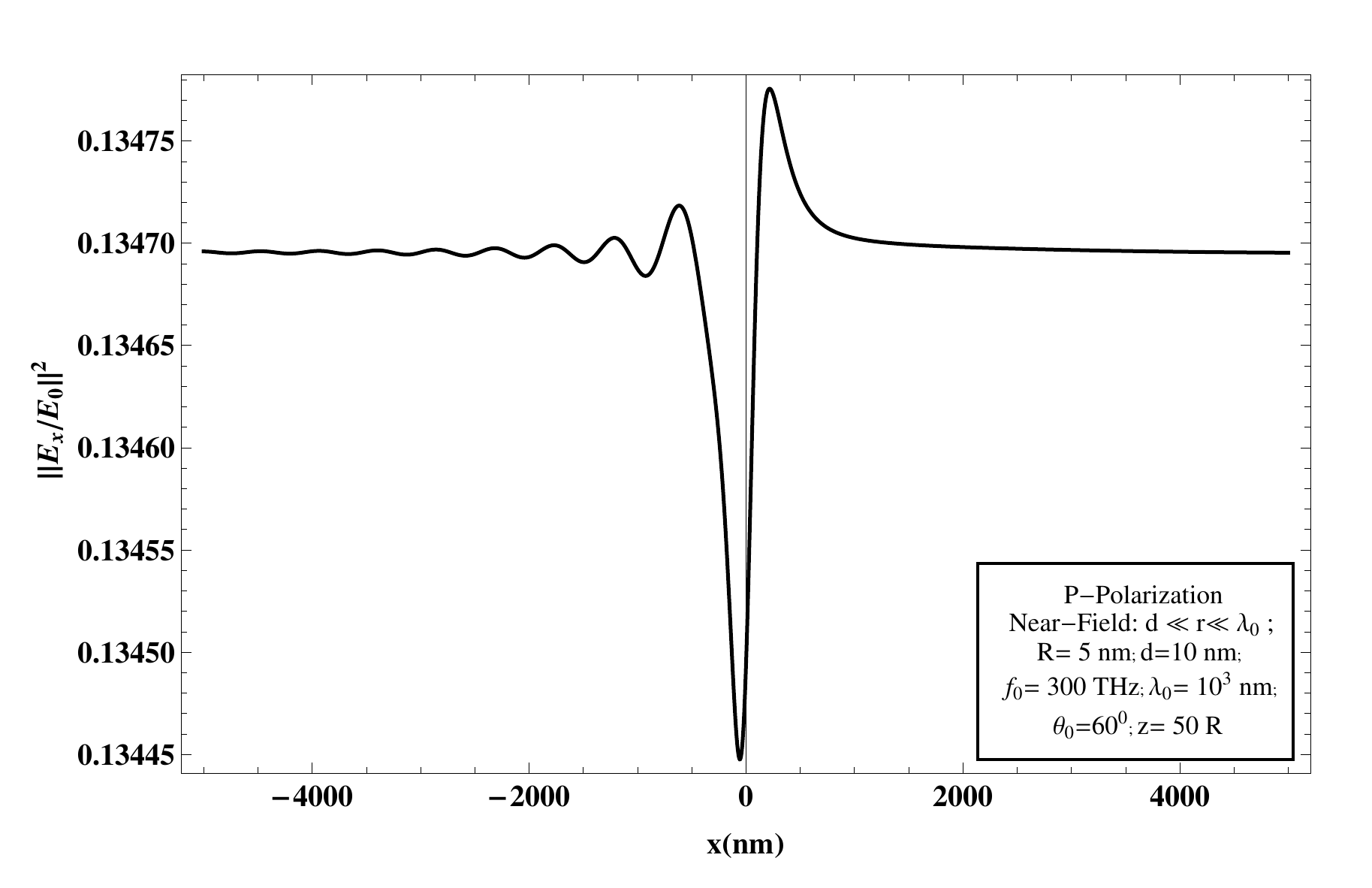}\\
(b)\\
 \includegraphics[width=9cm,height=7cm]{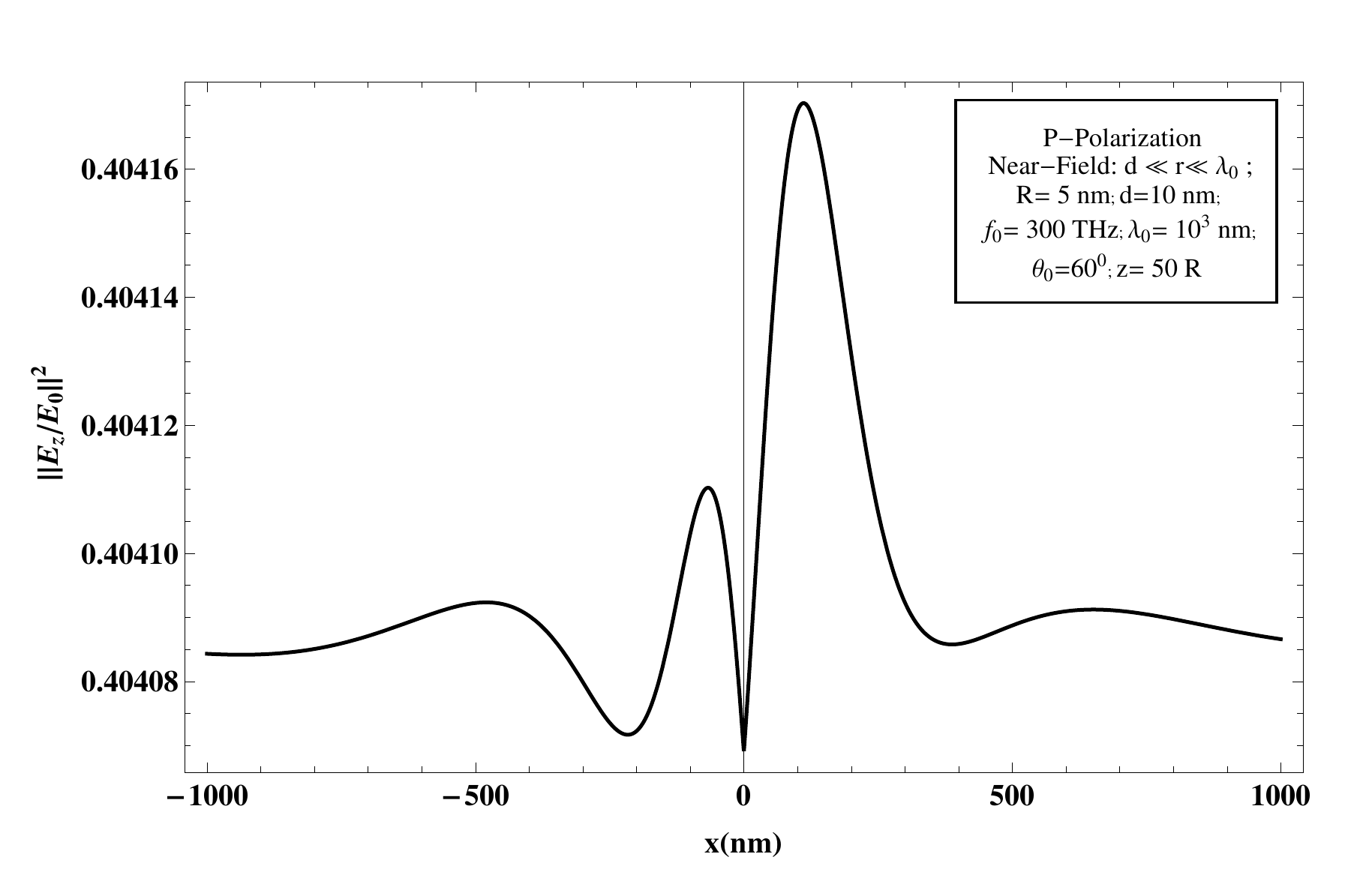}
\caption{$p$-polarization-Near-Field, $ z = 50\,R $; \textbf{$ \theta_{0}=60^{\circ}$}:
\newline (a) $\mid {E}_{x}(x,y,z;t)/{E}_{0}\mid^{2}$  and (b) $\mid {E}_{z}(x,y,z;t)/{E}_{0}\mid^{2}$ produced by a perforated
2D plasmonic layer of GaAs as a function of lateral distance $ r_{_{\parallel}}= x\,(y=0)$ from the aperture.}
\label{FIGNFT60ExEz}
\qquad
\end{figure}

\begin{figure}[h]
\centering
(a)\\
 \includegraphics[width=9cm,height=7cm]{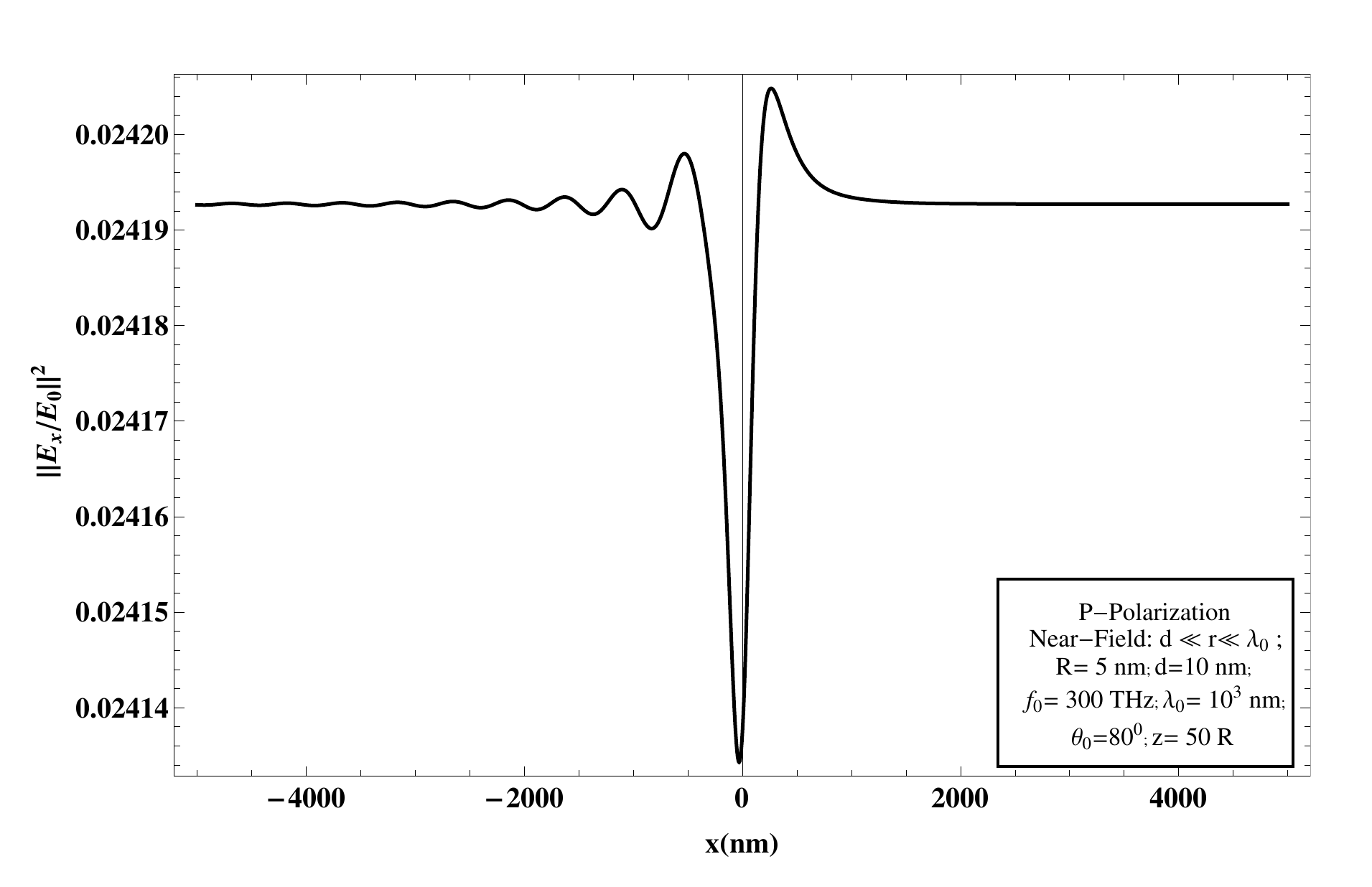}\\
(b)\\
 \includegraphics[width=9cm,height=7cm]{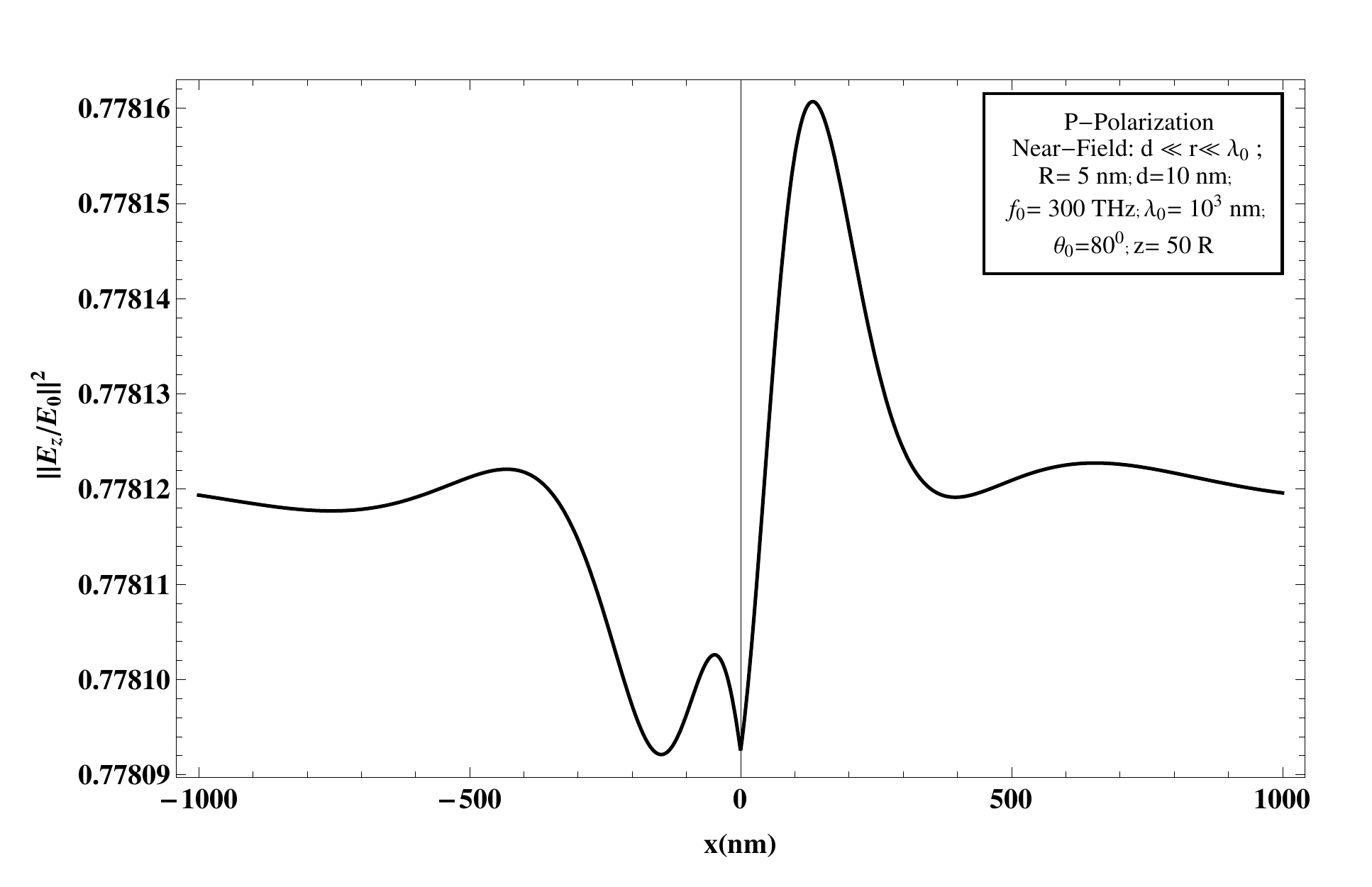}
\caption{$p$-polarization-Near-Field, $ z = 50\,R $; \textbf{$ \theta_{0}=80^{\circ}$}:
\newline (a) $\mid {E}_{x}(x,y,z;t)/{E}_{0}\mid^{2}$  and (b) $\mid {E}_{z}(x,y,z;t)/{E}_{0}\mid^{2}$ produced by a perforated
2D plasmonic layer of GaAs as a function of lateral distance $ r_{_{\parallel}}= x\,(y=0)$ from the aperture.}
\label{FIGNFT80ExEz}
\qquad
\end{figure}

\begin{figure}[h]
\centering
(a) \\
\includegraphics[width=9cm,height=6cm]{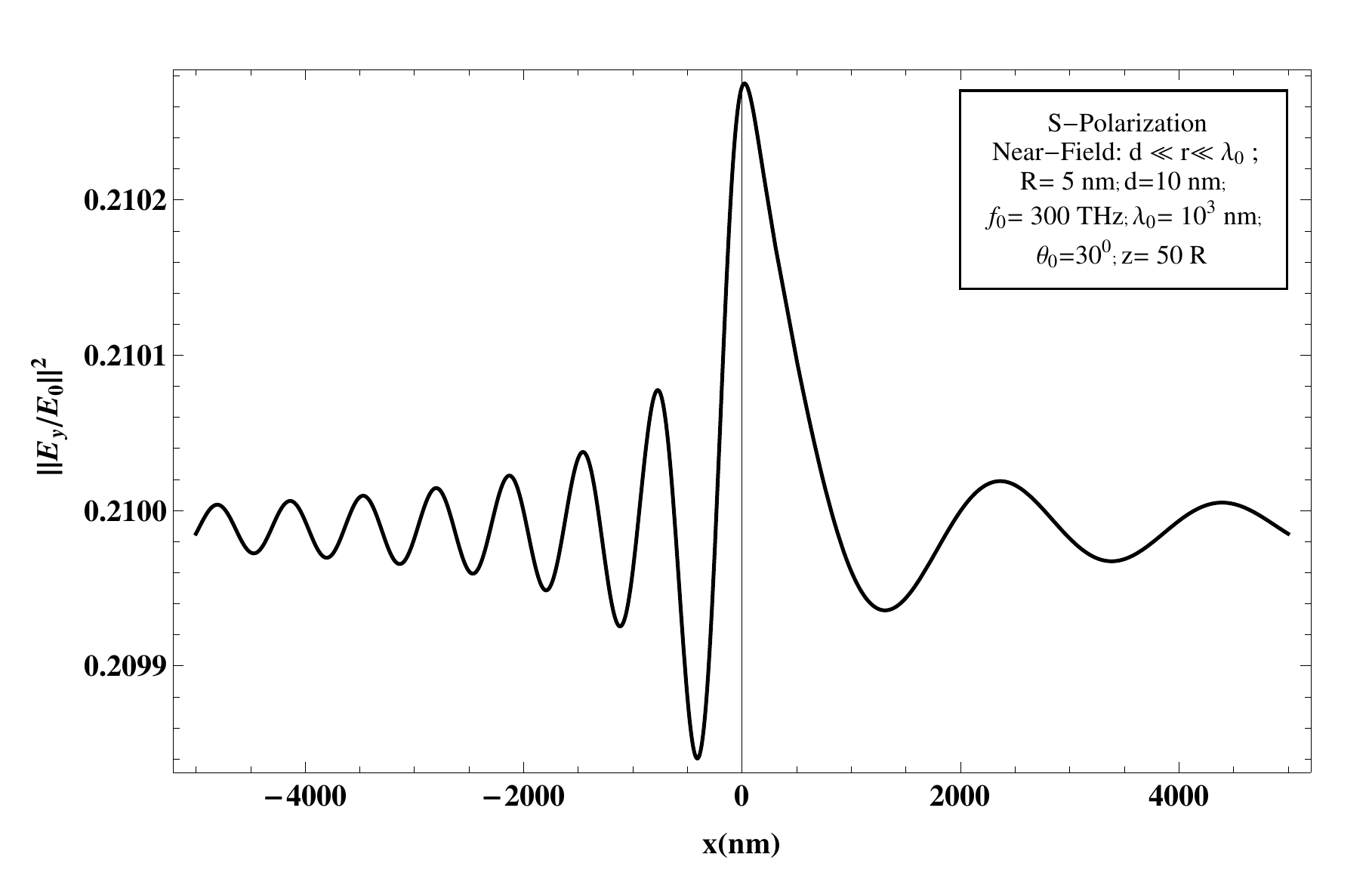}\\
(b)\\
 \includegraphics[width=9cm,height=6cm]{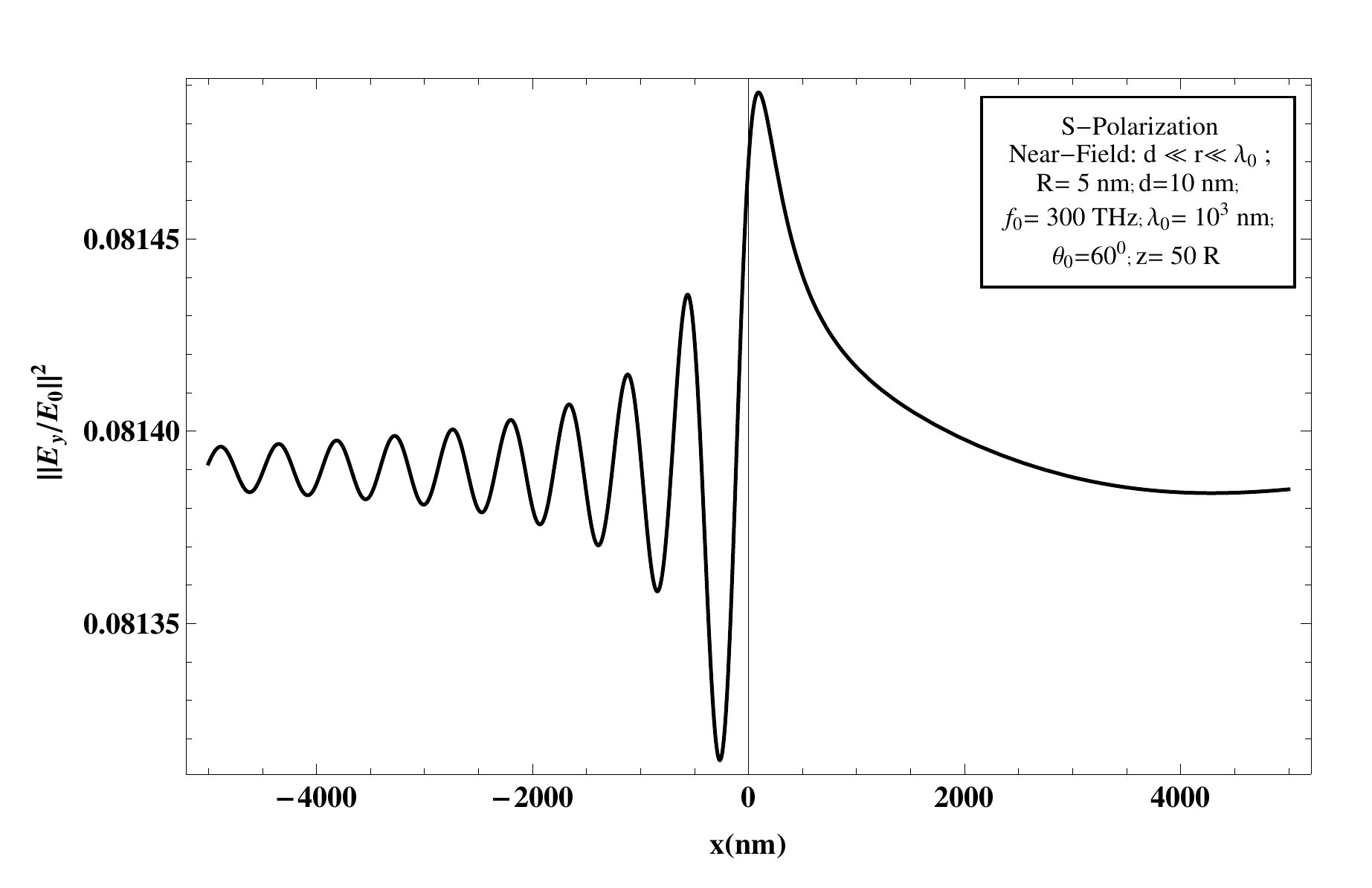}\\
(c)\\
\includegraphics[width=9cm,height=6cm]{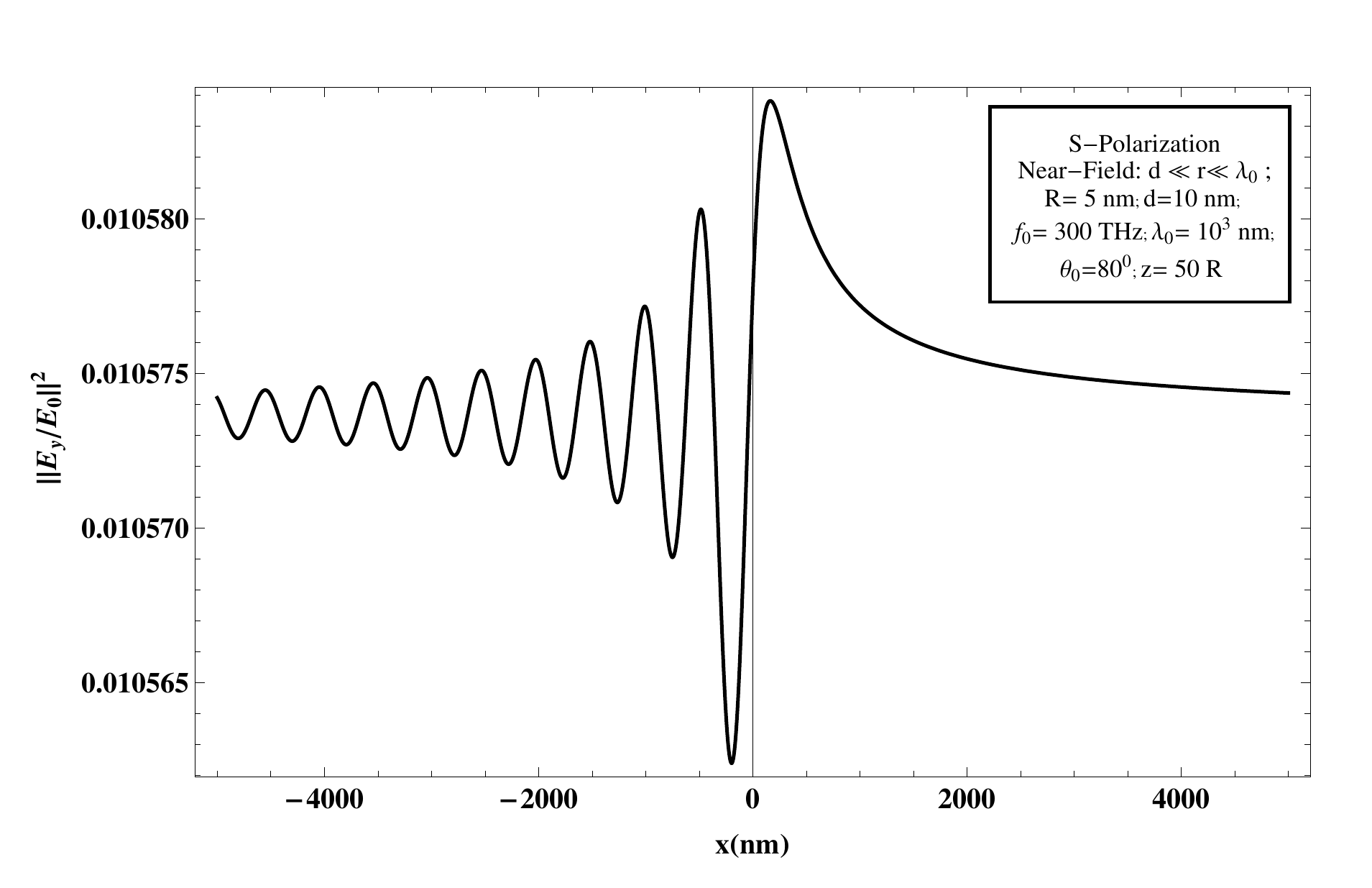}
\caption{$s$-polarization-Near-Field, $ z = 50\,R $ : $\mid {E}_{y}(x,y,z;t)/{E}_{0}\mid^{2}$ (a) $ \theta_{0}=30^{\circ}$, (b) $ \theta_{0}=60^{\circ}$
and (c) $ \theta_{0}=80^{\circ}$ produced by a perforated 2D plasmonic layer of GaAs as a function of lateral distance $ r_{_{\parallel}}= x\,(y=0)$ from the aperture.}
\label{FIGNFT306080Ey}
\qquad
\end{figure}

\begin{figure}[h]
\centering
(a)\\
 \includegraphics[width=9cm,height=7cm]{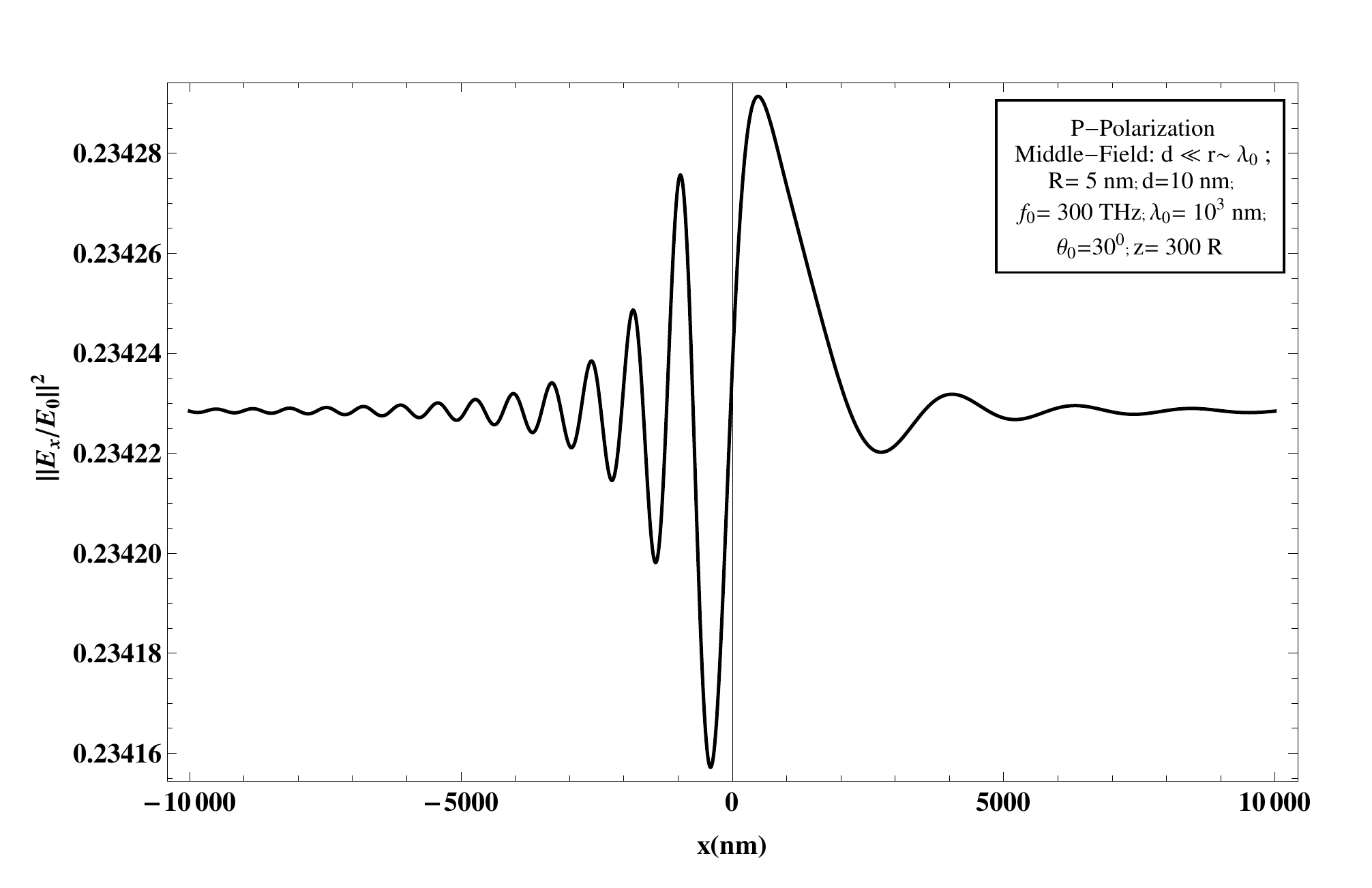}\\
(b)\\
 \includegraphics[width=9cm,height=7cm]{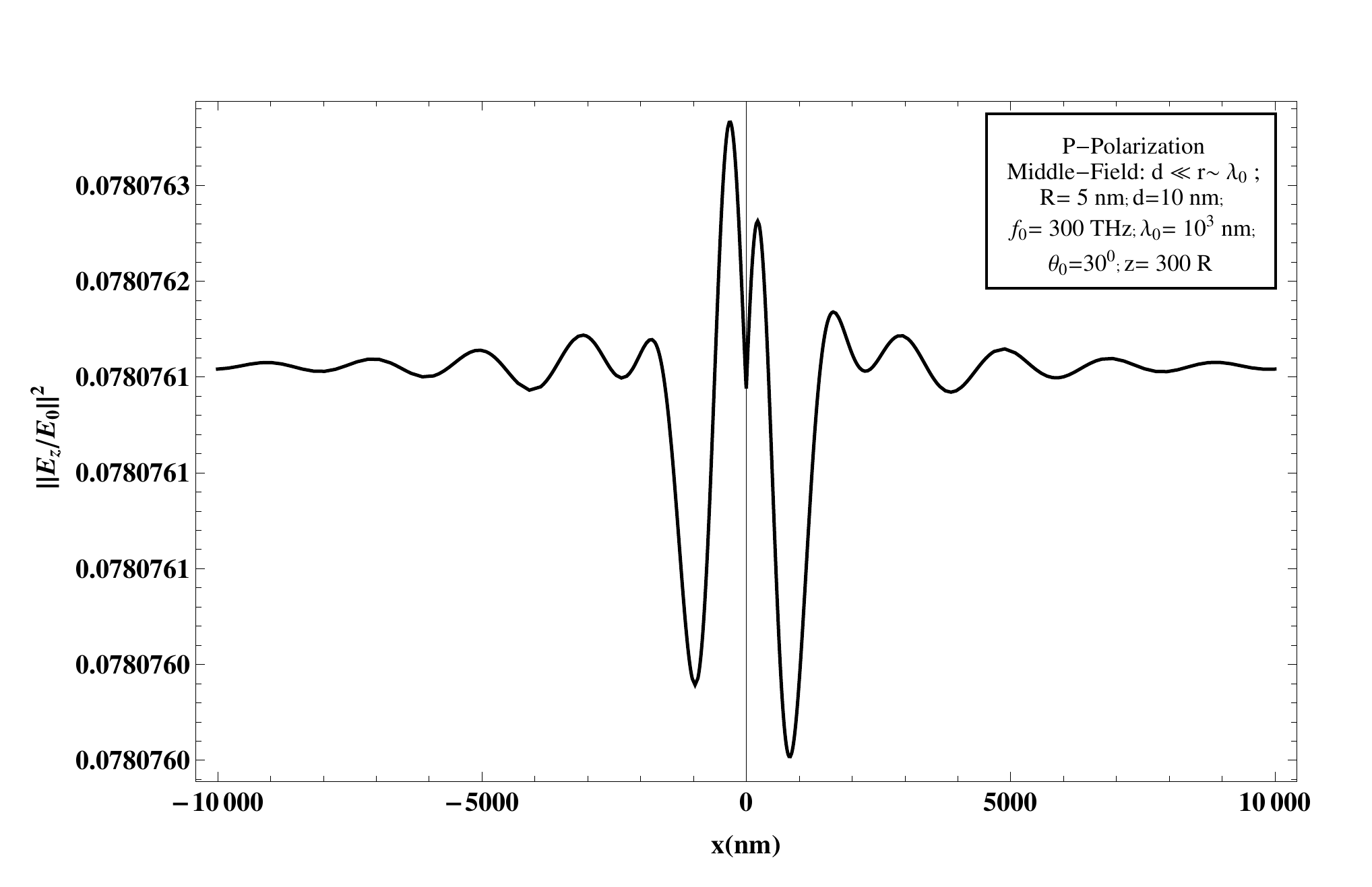}
\caption{$ p $ - polarization-Middle-Field, $ z = 300\,R $; $ \theta_{0}=30^{\circ}$:
\newline
(a) $\mid {E}_{x}(x,y,z;t)/{E}_{0}\mid^{2}$  and (b) $\mid {E}_{z}(x,y,z;t)/{E}_{0}\mid^{2}$ produced by a perforated 2D plasmonic layer of GaAs as a function of lateral distance $ r_{_{\parallel}}= x\,(y=0)$ from the aperture.}
\label{FIGMFT30ExEz}
\qquad
\end{figure}

\begin{figure}[h]
\centering
(a)\\
 \includegraphics[width=9cm,height=7cm]{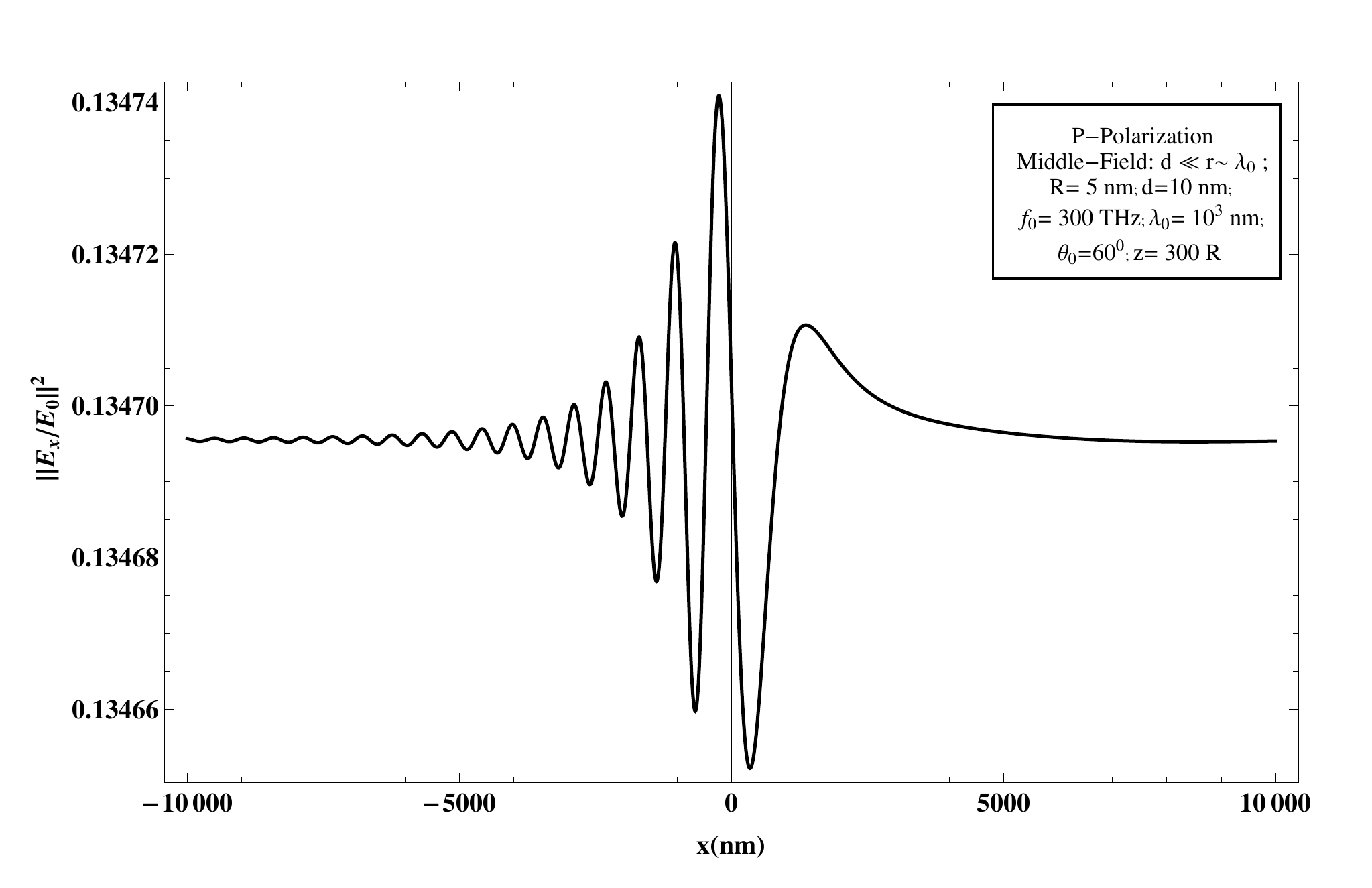}\\
(b)\\
 \includegraphics[width=9cm,height=7cm]{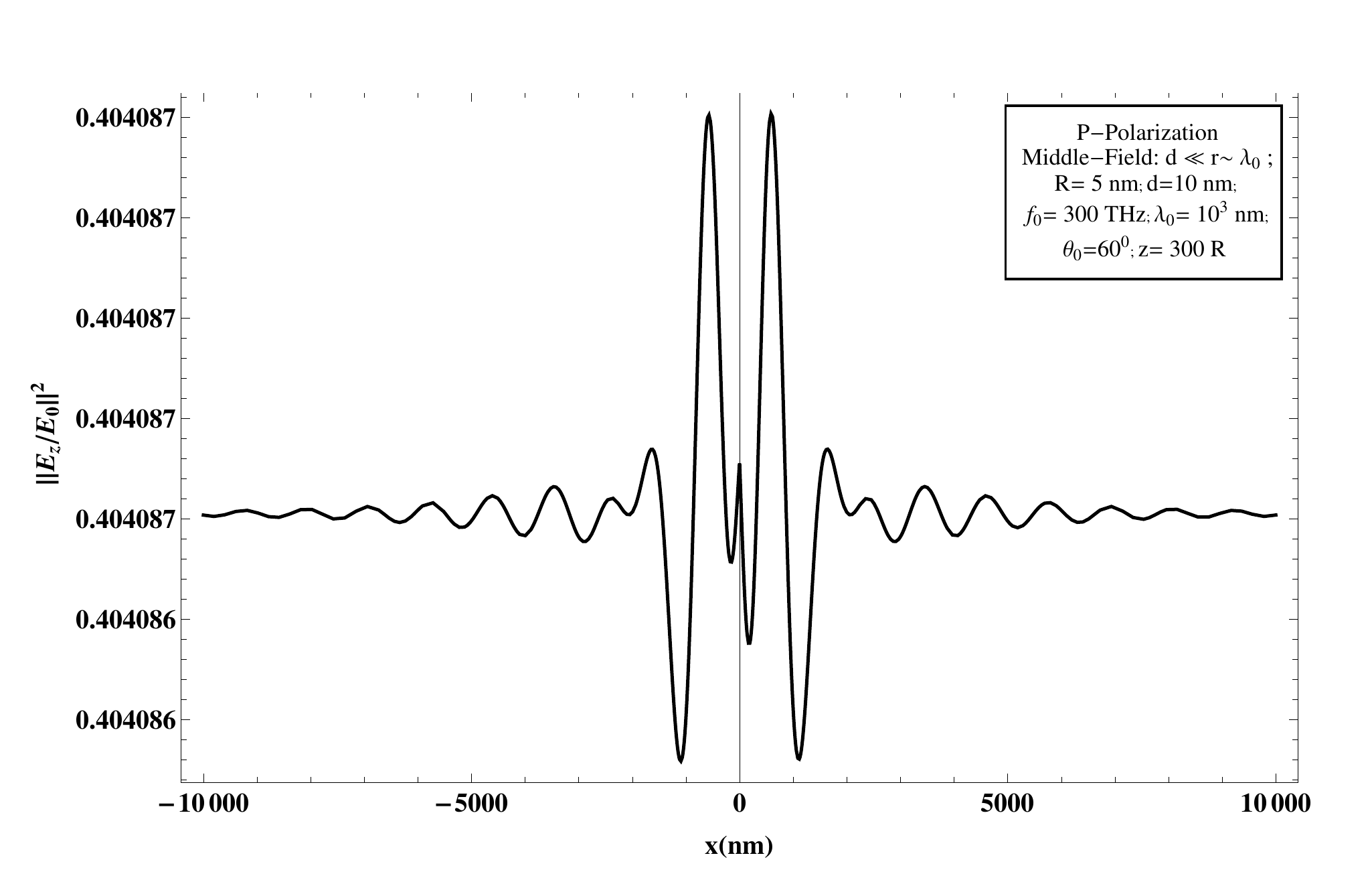}
\caption{$ p $ - polarization-Middle-Field, $ z = 300\,R $; $ \theta_{0}=60^{\circ}$:
\newline
(a) $\mid {E}_{x}(x,y,z;t)/{E}_{0}\mid^{2}$  and (b) $\mid {E}_{z}(x,y,z;t)/{E}_{0}\mid^{2}$ produced by a perforated 2D plasmonic layer of GaAs as a function of lateral distance $ r_{_{\parallel}}= x\,(y=0)$ from the aperture.}
\label{FIGMFT60ExEz}
\qquad
\end{figure}

\begin{figure}[h]
\centering
(a)\\
 \includegraphics[width=9cm,height=7cm]{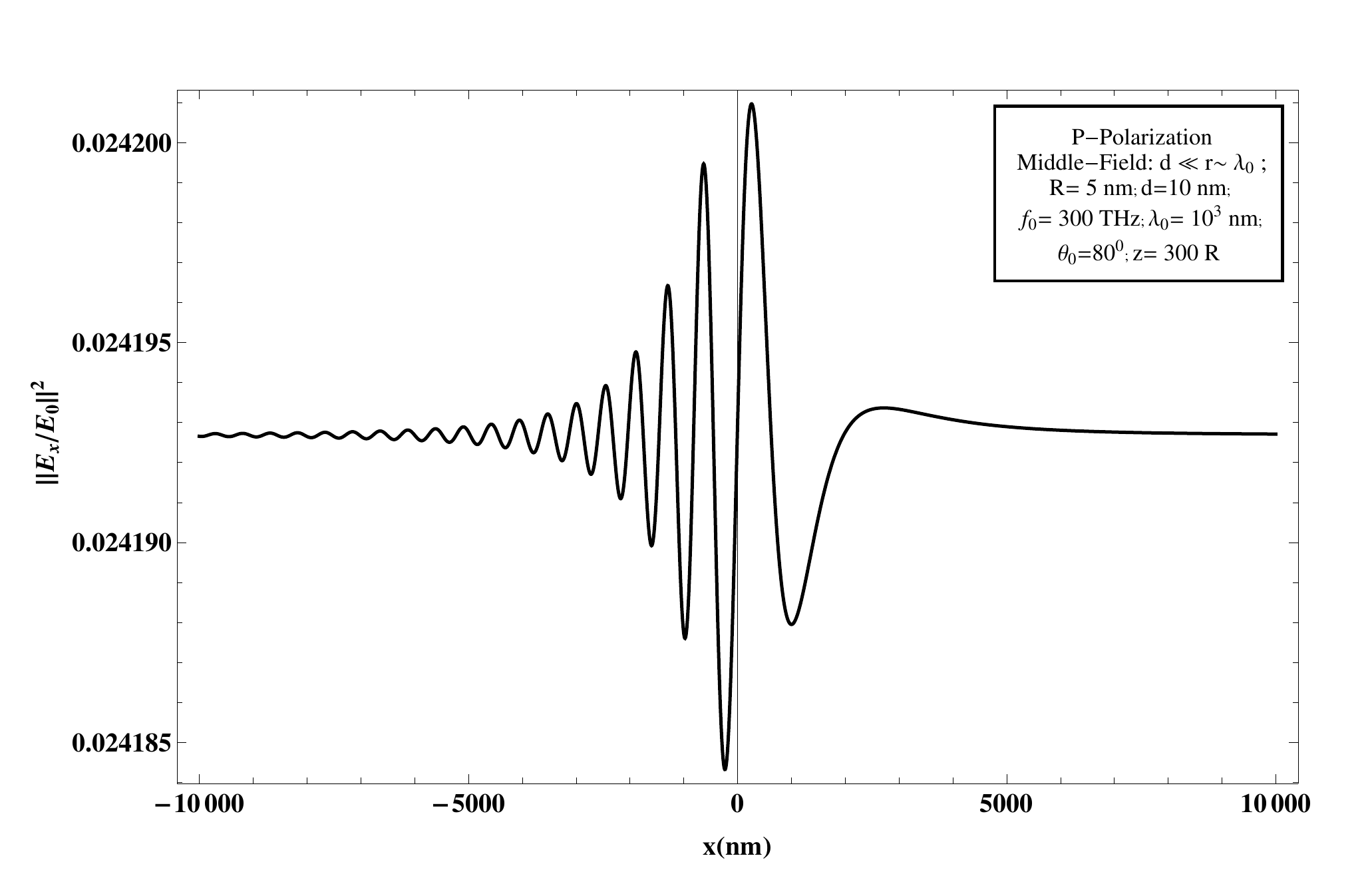}\\
(b)\\
 \includegraphics[width=9cm,height=7cm]{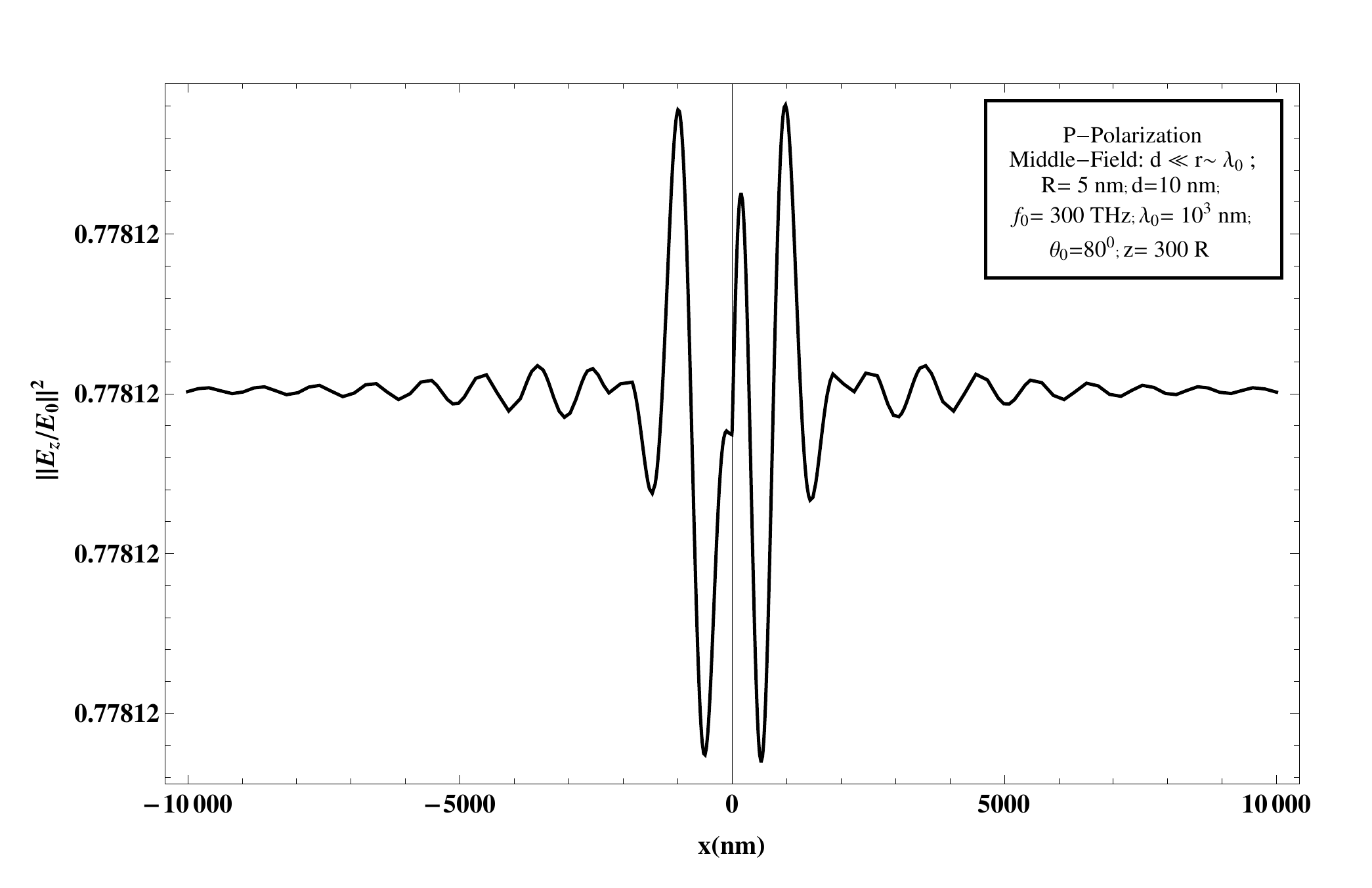}
\caption{$ p $ - polarization-Middle-Field, $ z = 300\,R $; $ \theta_{0}=80^{\circ}$:
\newline
(a) $\mid {E}_{x}(x,y,z;t)/{E}_{0}\mid^{2}$  and (b) $\mid {E}_{z}(x,y,z;t)/{E}_{0}\mid^{2}$ produced by a perforated
2D plasmonic layer of GaAs as a function of lateral distance $ r_{_{\parallel}}= x\,(y=0)$ from the aperture.}
\label{FIGMFT80ExEz}
\qquad
\end{figure}

\begin{figure}[h]
\centering
(a)\\
 \includegraphics[width=9cm,height=6cm]{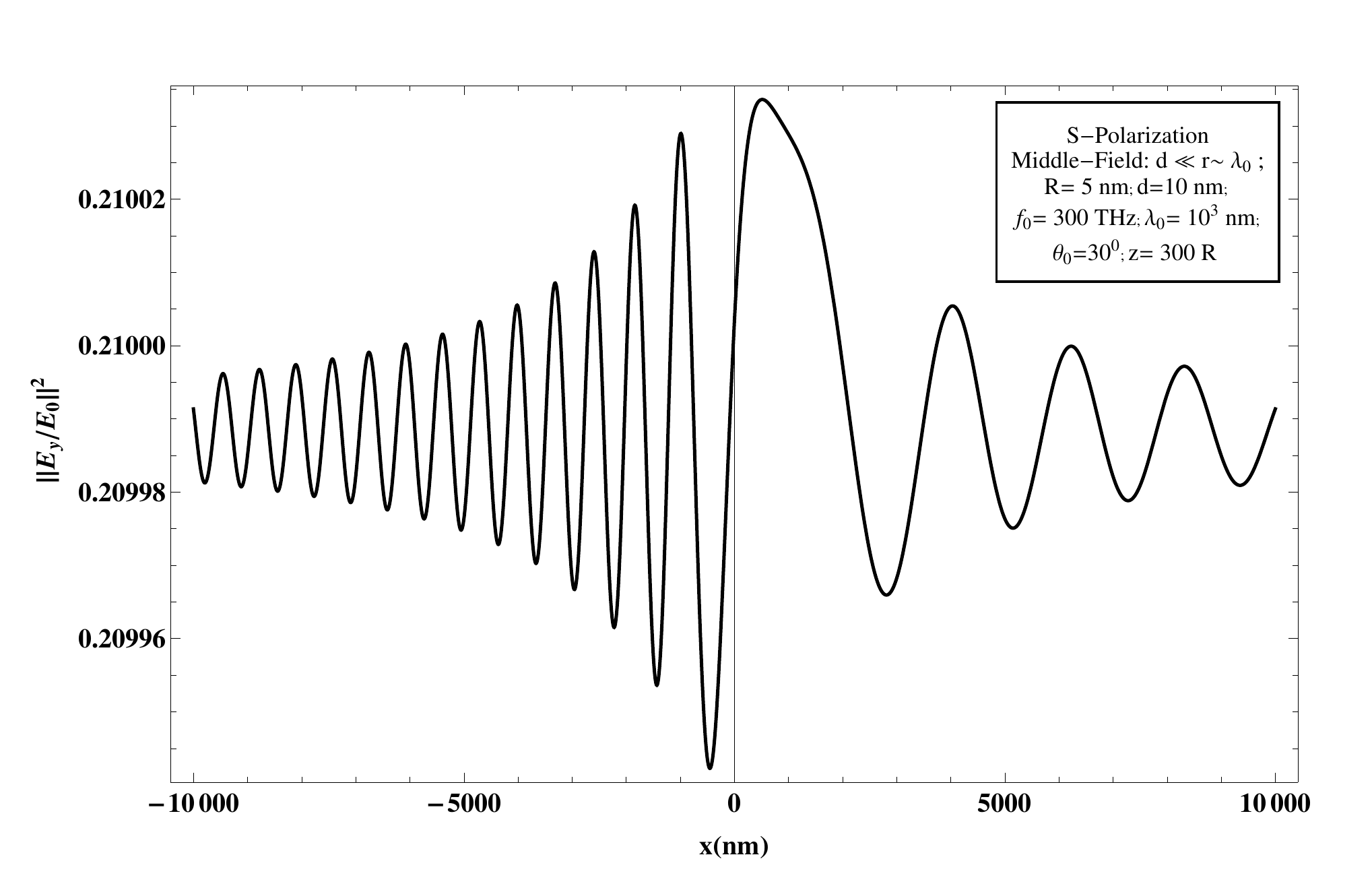}\\
(b)\\
 \includegraphics[width=9cm,height=6cm]{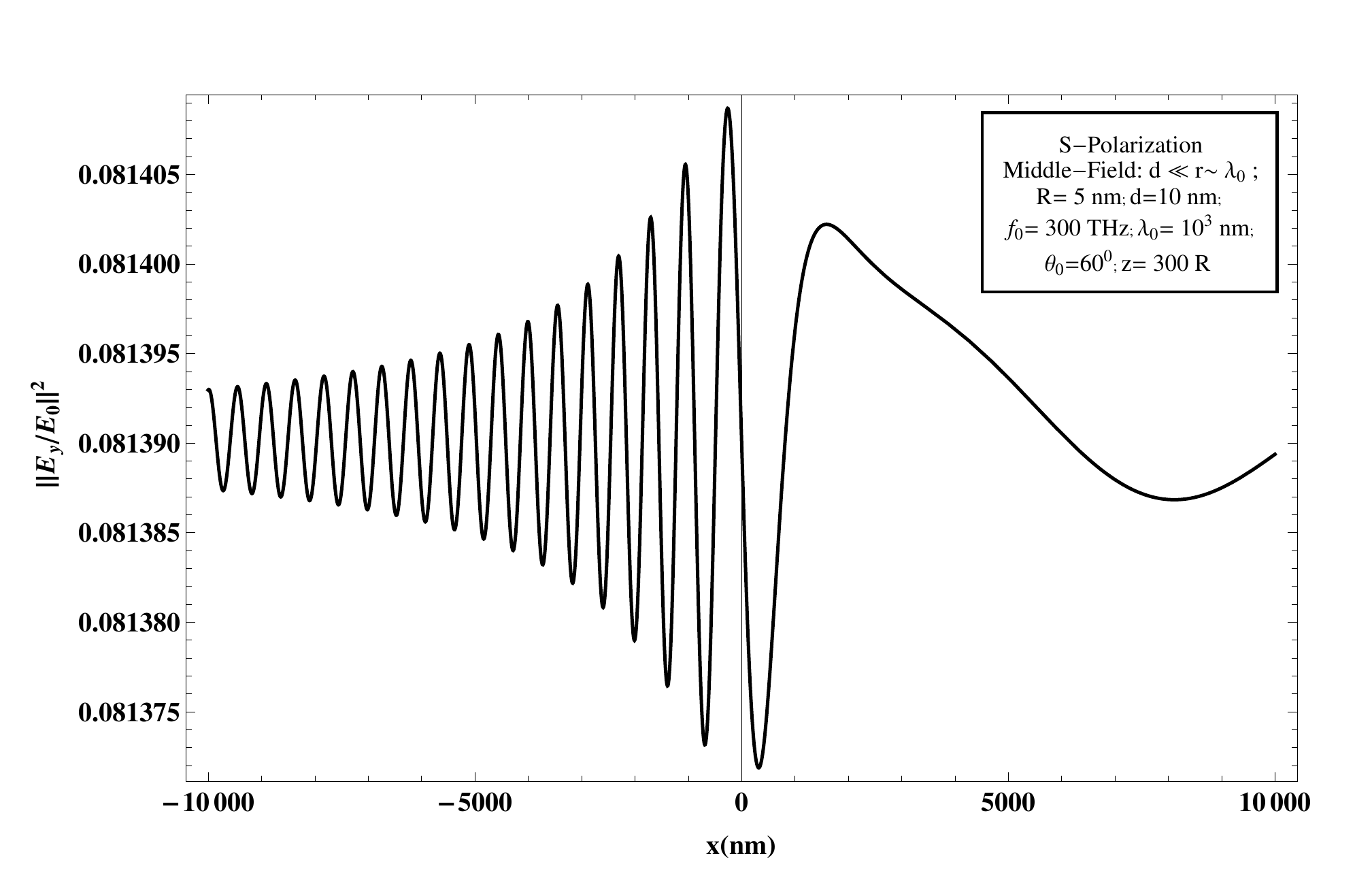}\\
(c)\\
 \includegraphics[width=9cm,height=6cm]{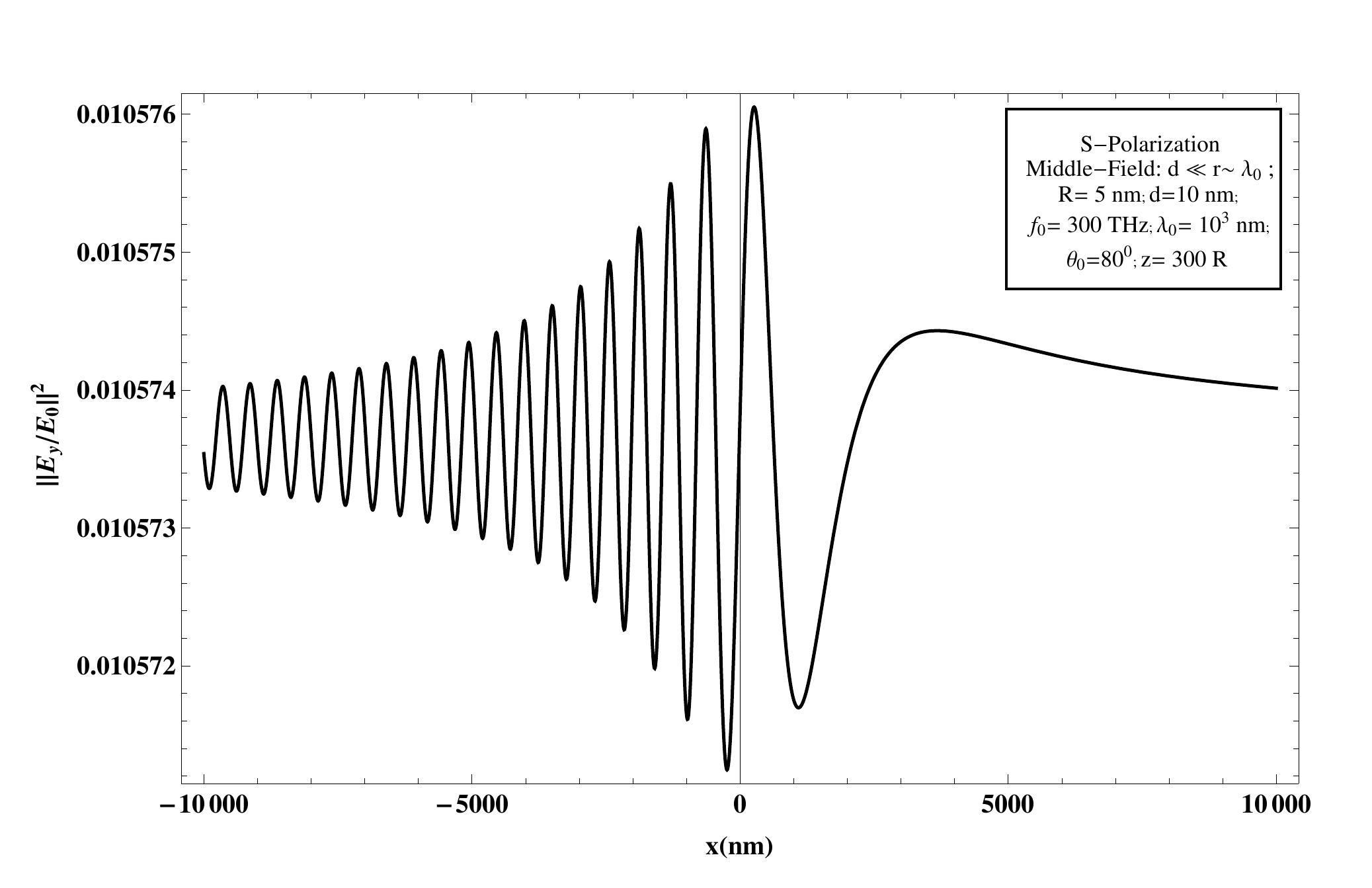}
\caption{ $ s $ - polarization-Middle-Field, $ z = 300\,R $; $\mid {E}_{y}(x,y,z;t)/{E}_{0}\mid^{2}$:
(a) $ \theta_{0}=30^{\circ}$ , (b) $ \theta_{0}=60^{\circ}$
and (c) $ \theta_{0}=80^{\circ}$ produced by a perforated
2D plasmonic layer of GaAs as a function of lateral distance $ r_{_{\parallel}}= x\,(y=0)$ from the aperture.}
\label{FIGMFT306080Ey}
\qquad
\end{figure}

\begin{figure}[h]
\centering
(a)\\
 \includegraphics[width=9cm,height=7cm]{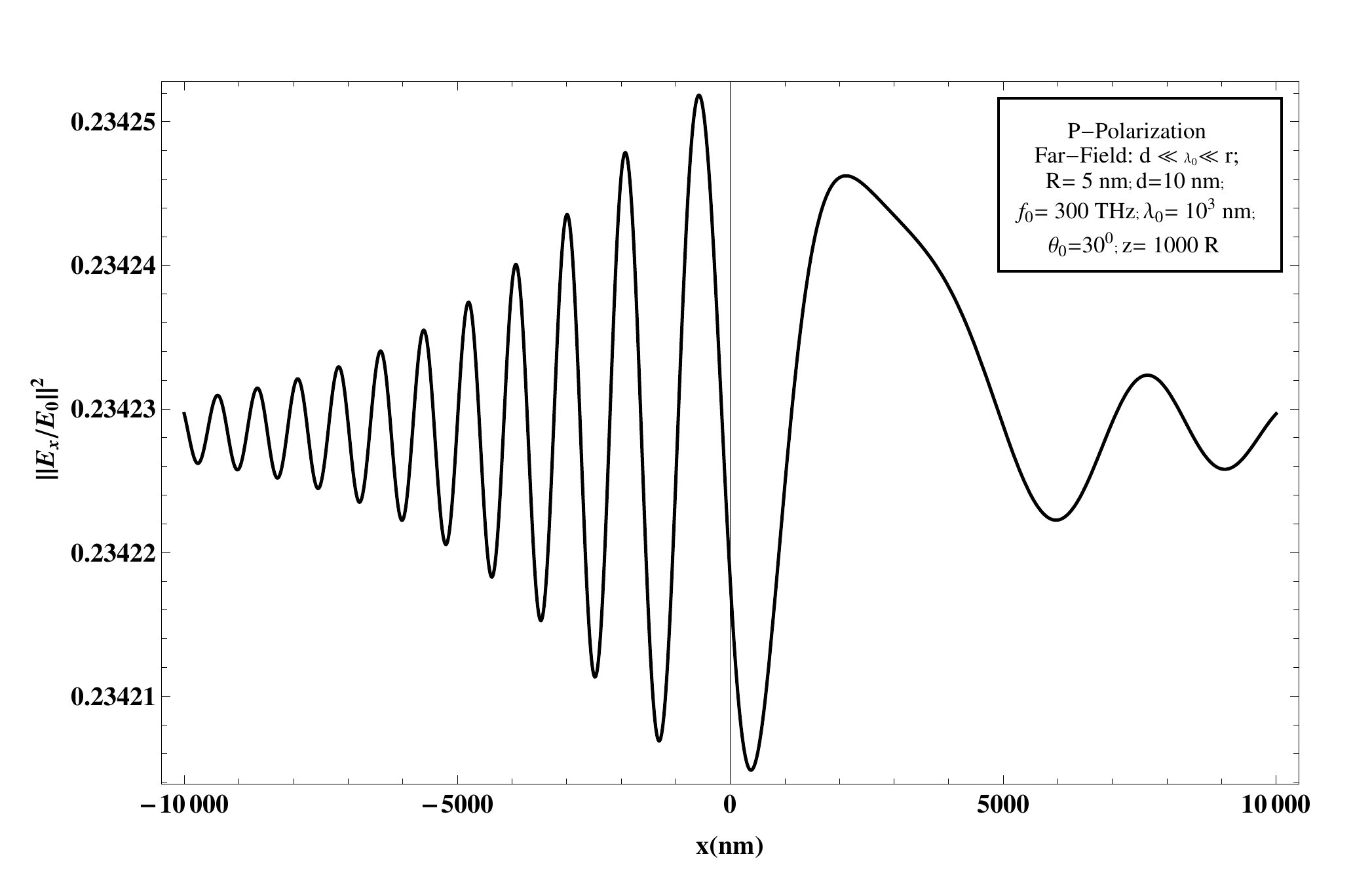}\\
(b)\\
 \includegraphics[width=9cm,height=7cm]{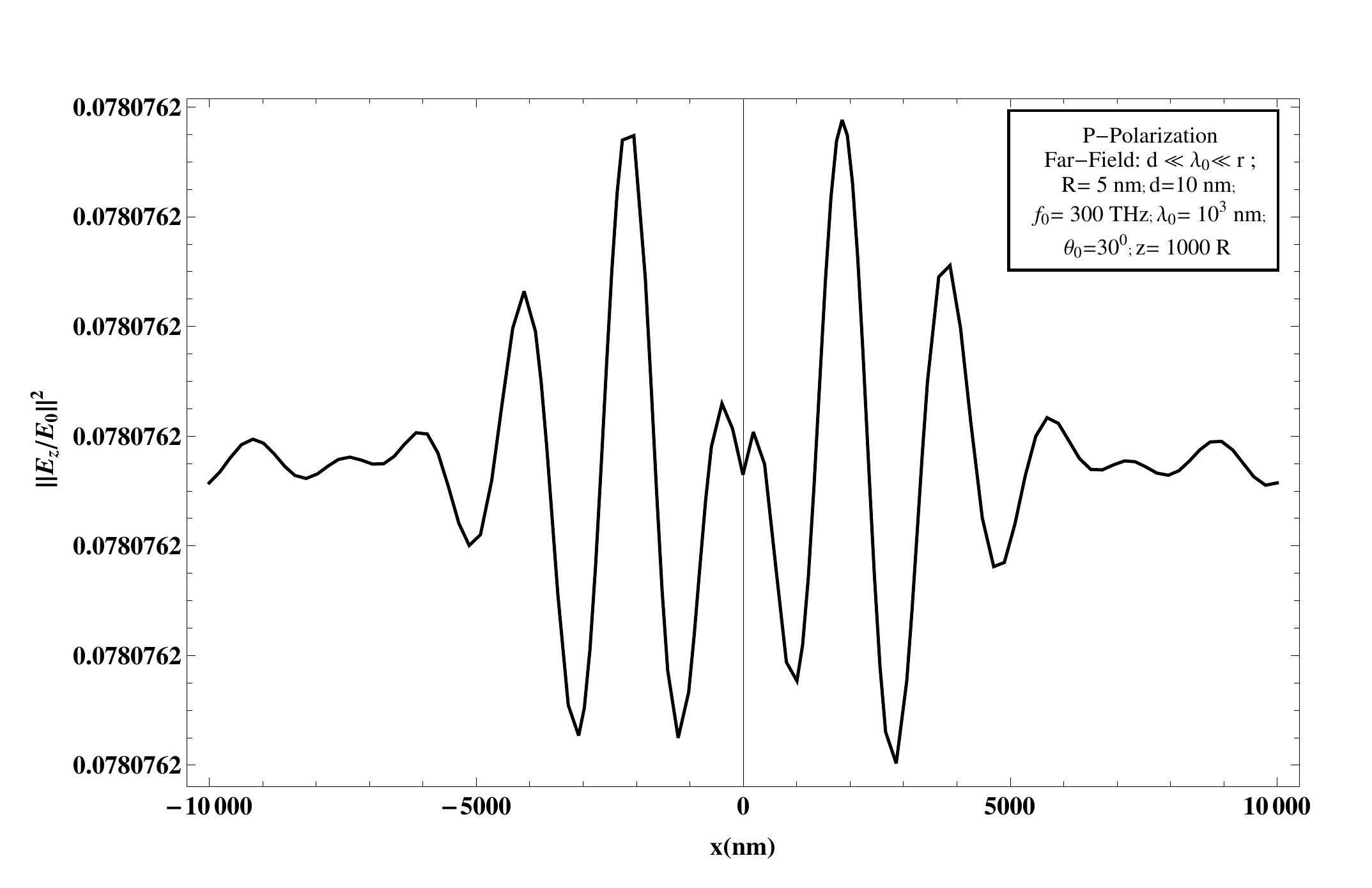}
\caption{$ p $ - polarization-Far-Field, $ z = 1000\,R $;$ \theta_{0}=30^{\circ}$:
\newline
(a) $\mid {E}_{x}(x,y,z;t)/{E}_{0}\mid^{2}$  and (b) $\mid {E}_{z}(x,y,z;t)/{E}_{0}\mid^{2}$ produced by a perforated
2D plasmonic layer of GaAs as a function of lateral distance $ r_{_{\parallel}}= x\,(y=0)$ from the aperture.}
\label{FIGFFT30ExEz}
\qquad
\end{figure}

\begin{figure}[h]
\centering
(a)\\
 \includegraphics[width=9cm,height=7cm]{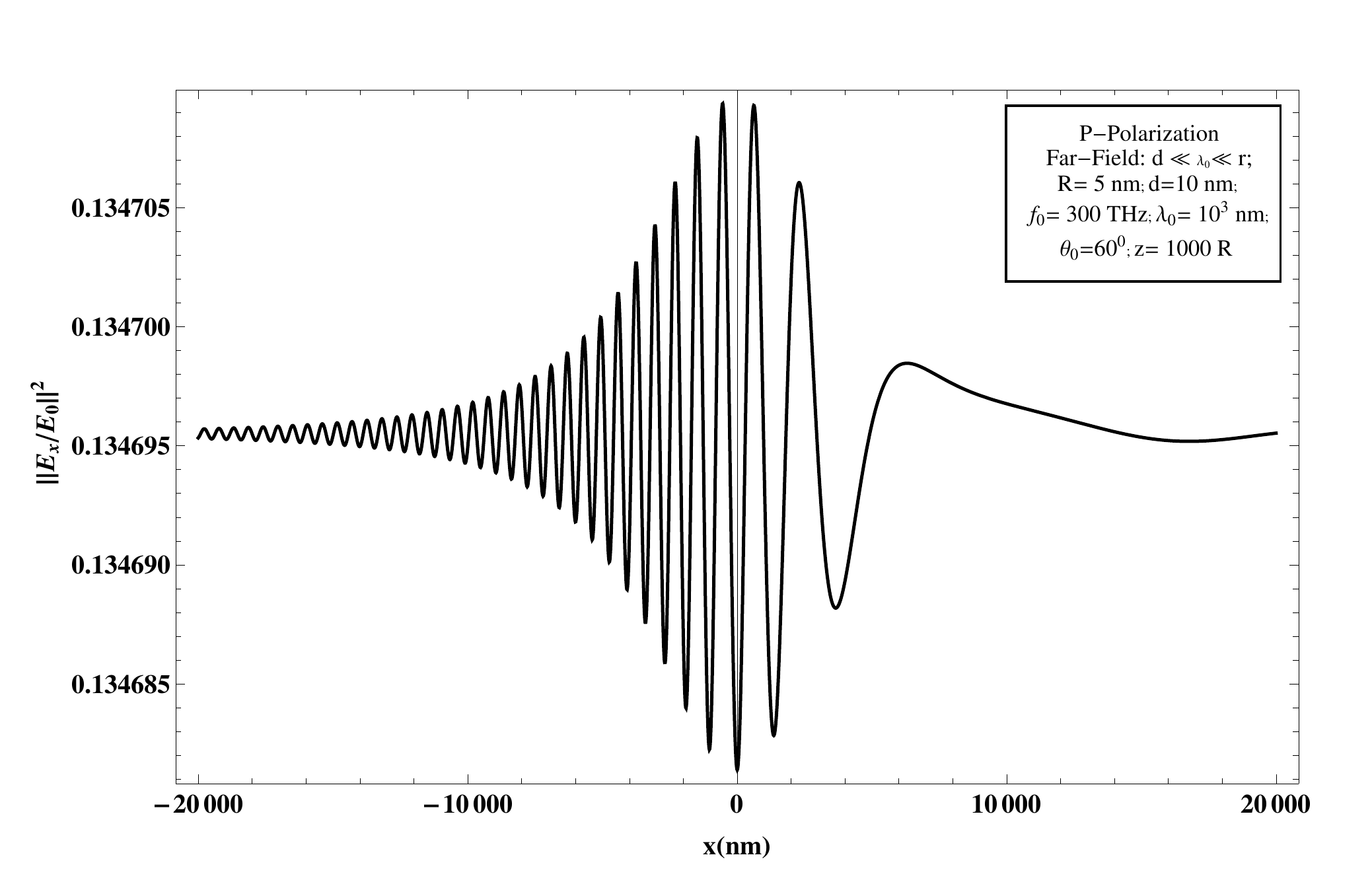}\\
(b)\\
 \includegraphics[width=9cm,height=7cm]{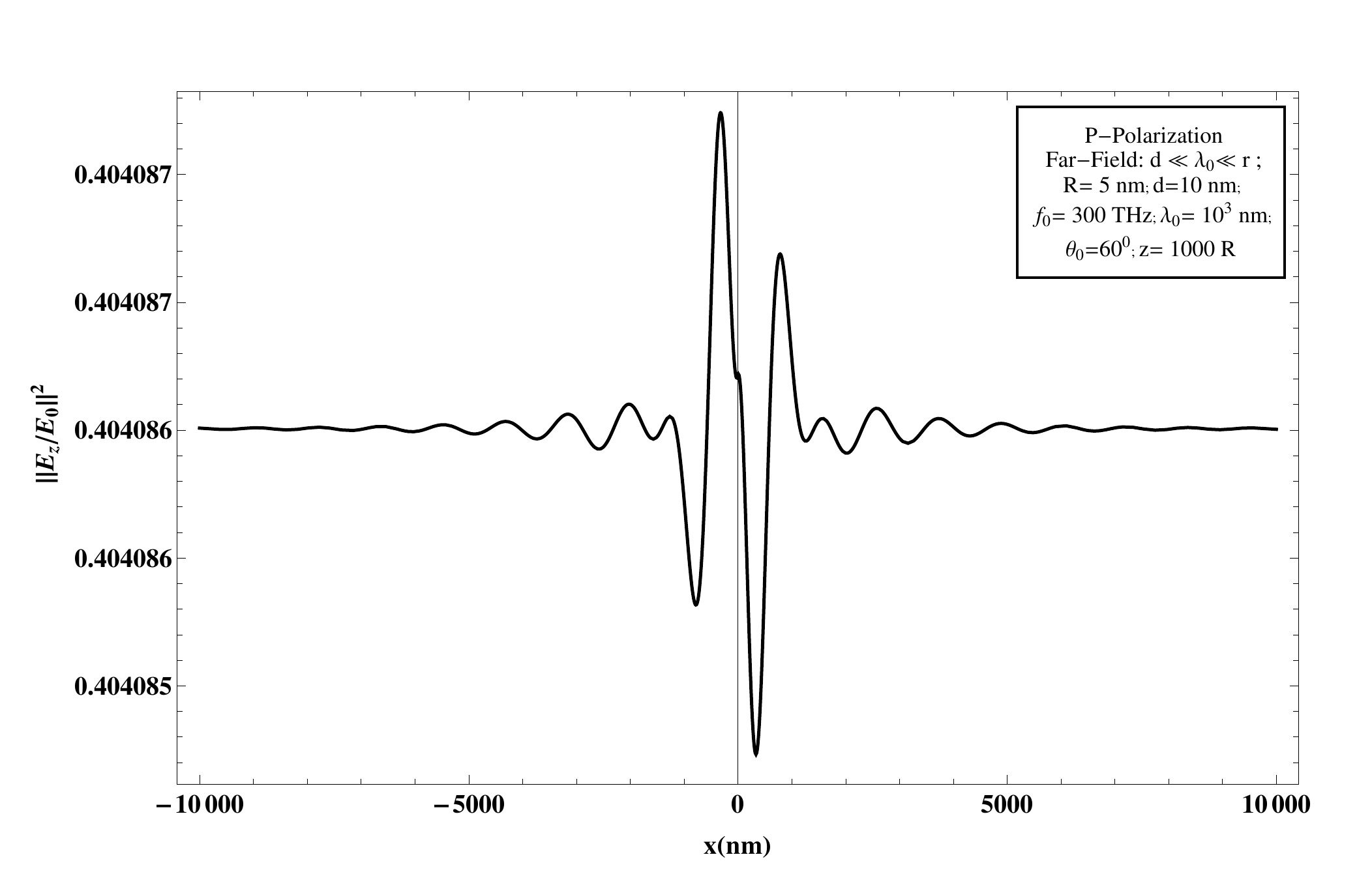}
\caption{$ p $ - polarization-Far-Field, $ z = 1000\,R $;$ \theta_{0}=60^{\circ}$:
\newline
(a) $\mid {E}_{x}(x,y,z;t)/{E}_{0}\mid^{2}$  and (b) $\mid {E}_{z}(x,y,z;t)/{E}_{0}\mid^{2}$ produced by a perforated
2D plasmonic layer of GaAs as a function of lateral distance $ r_{_{\parallel}}= x\,(y=0)$ from the aperture.}
\label{FIGFFT60ExEz}
\qquad
\end{figure}

\begin{figure}[h]
\centering
(a)\\
 \includegraphics[width=9cm,height=7cm]{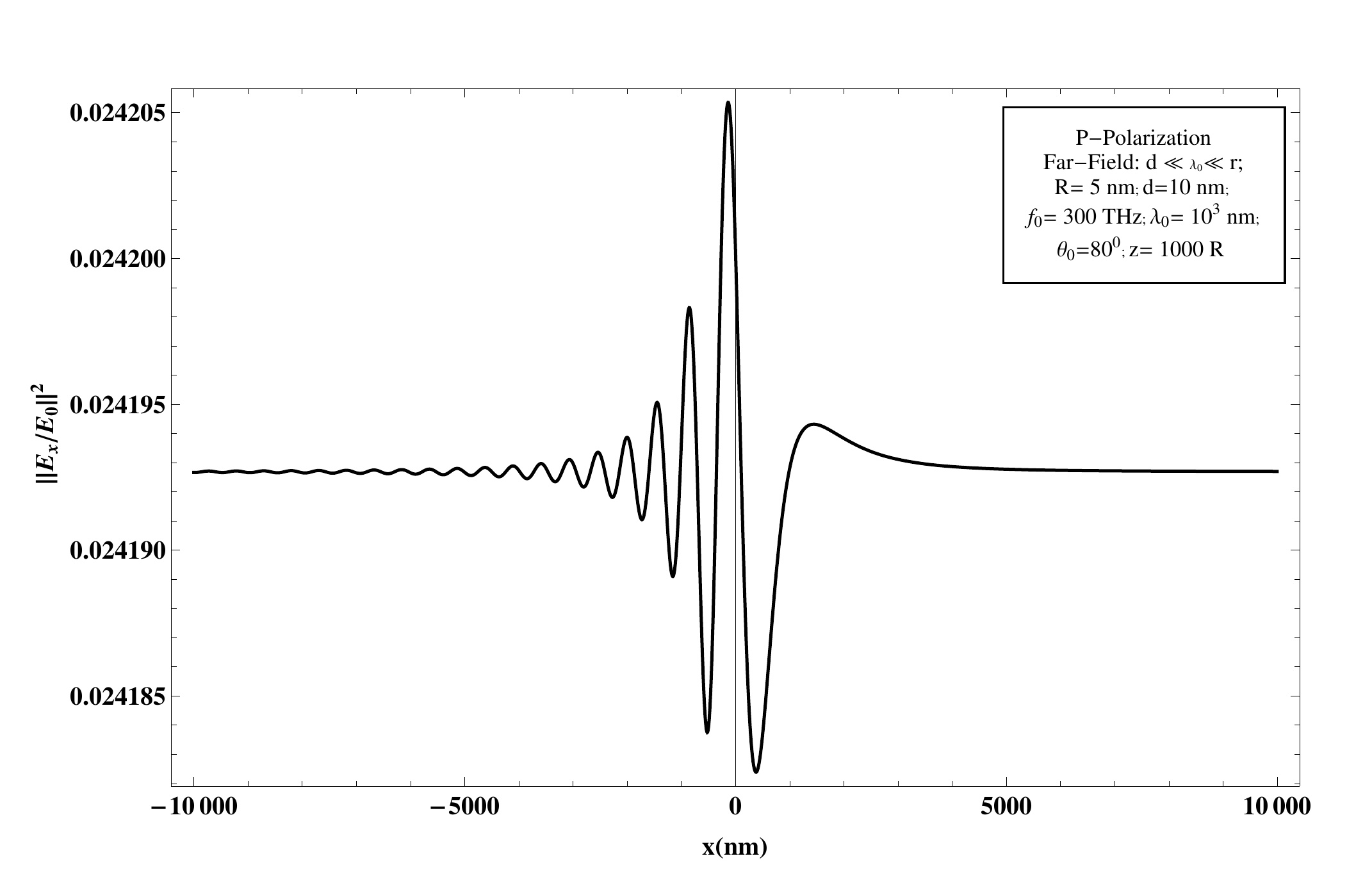}\\
(b)\\
 \includegraphics[width=9cm,height=7cm]{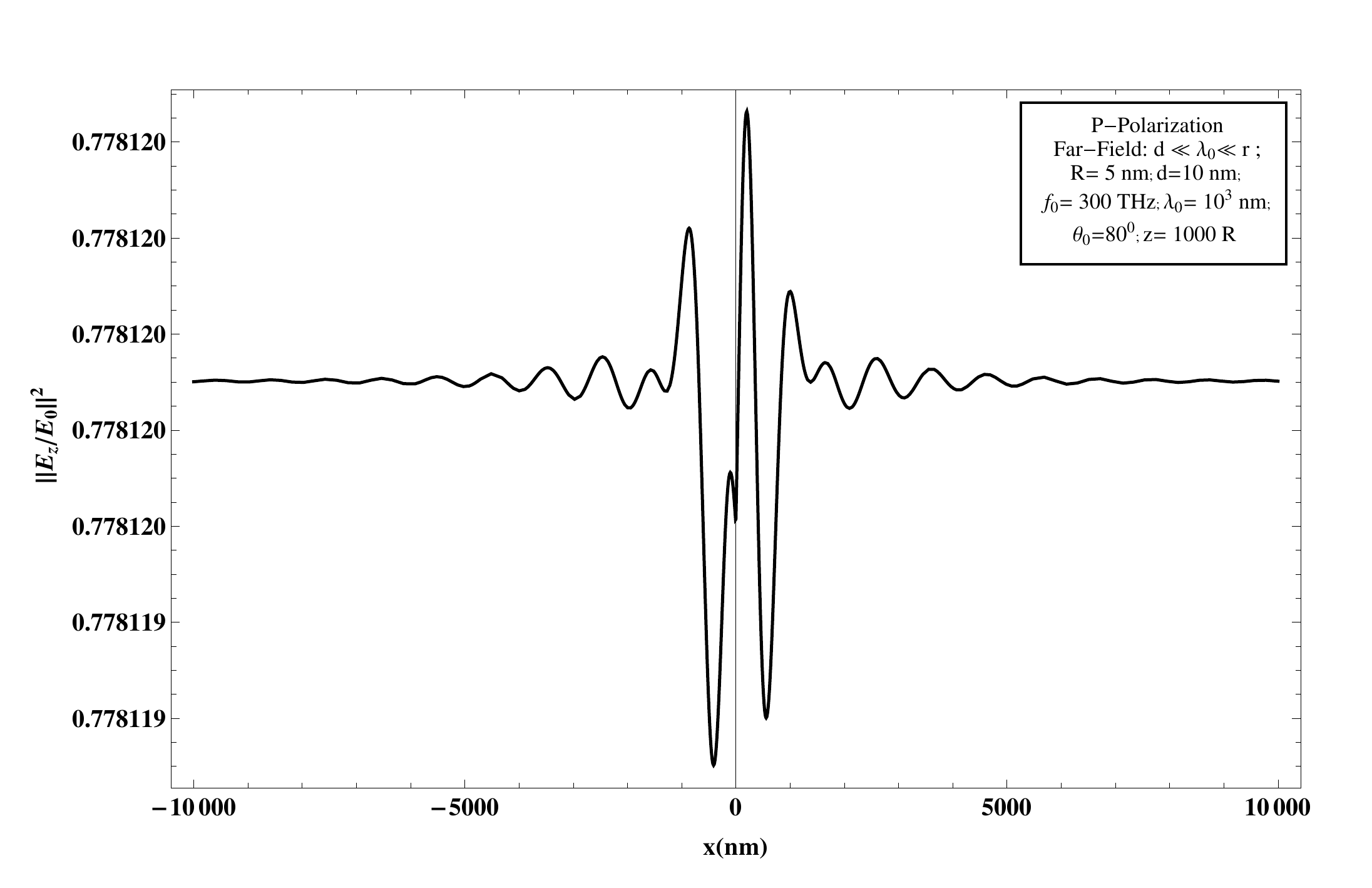}
\caption{$ p $ - polarization-Far-Field, $ z = 1000\,R $;$ \theta_{0}=80^{\circ}$:
\newline
(a) $\mid {E}_{x}(x,y,z;t)/{E}_{0}\mid^{2}$  and (b) $\mid {E}_{z}(x,y,z;t)/{E}_{0}\mid^{2}$ produced by a perforated
2D plasmonic layer of GaAs as a function of lateral distance $ r_{_{\parallel}}= x\,(y=0)$ from the aperture.}
\label{FIGFFT80ExEz}
\qquad
\end{figure}

\begin{figure}[h]
\centering
(a)\\
\includegraphics[width=9cm,height=6cm]{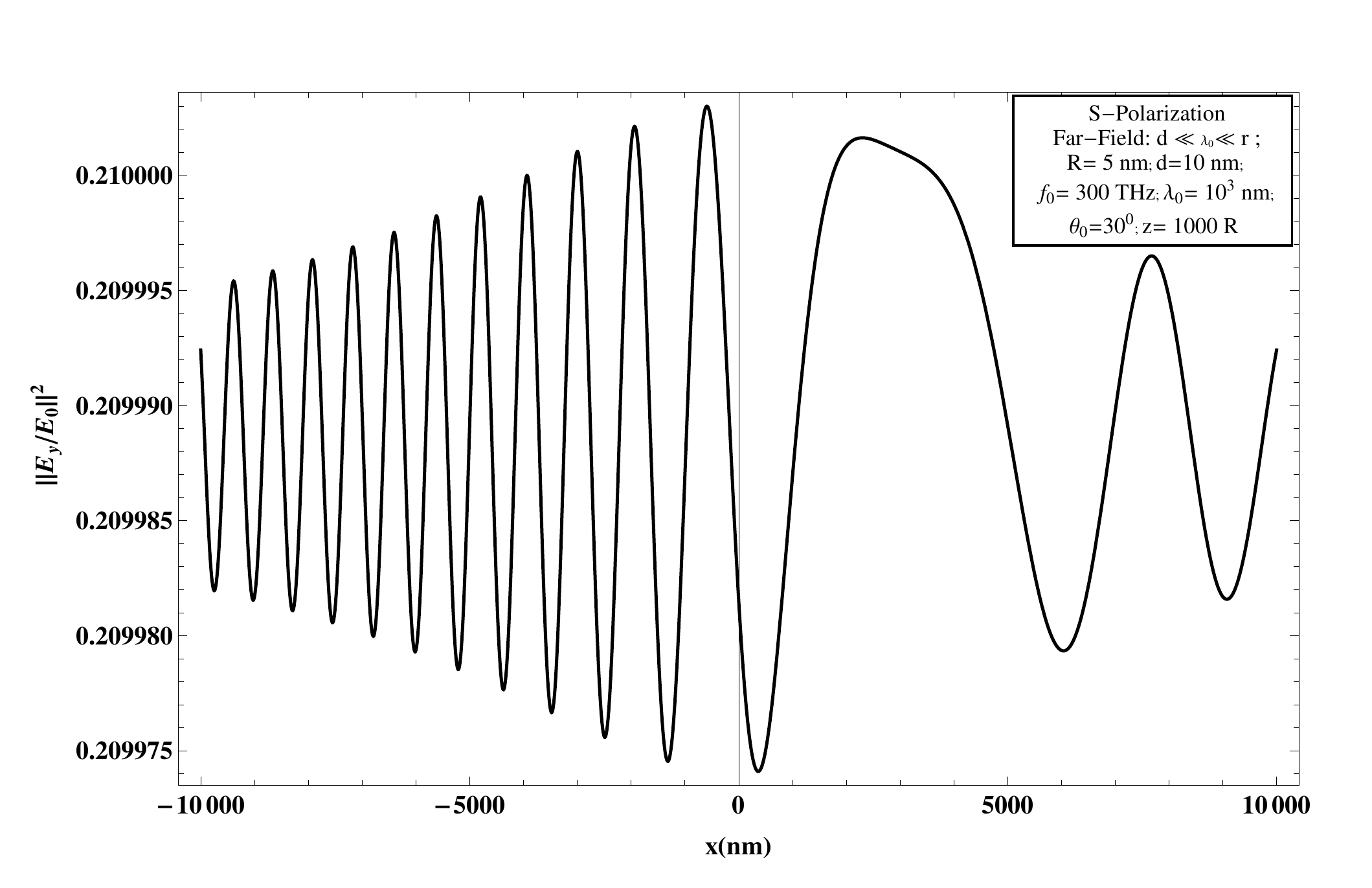}\\
(b)\\
\includegraphics[width=9cm,height=6cm]{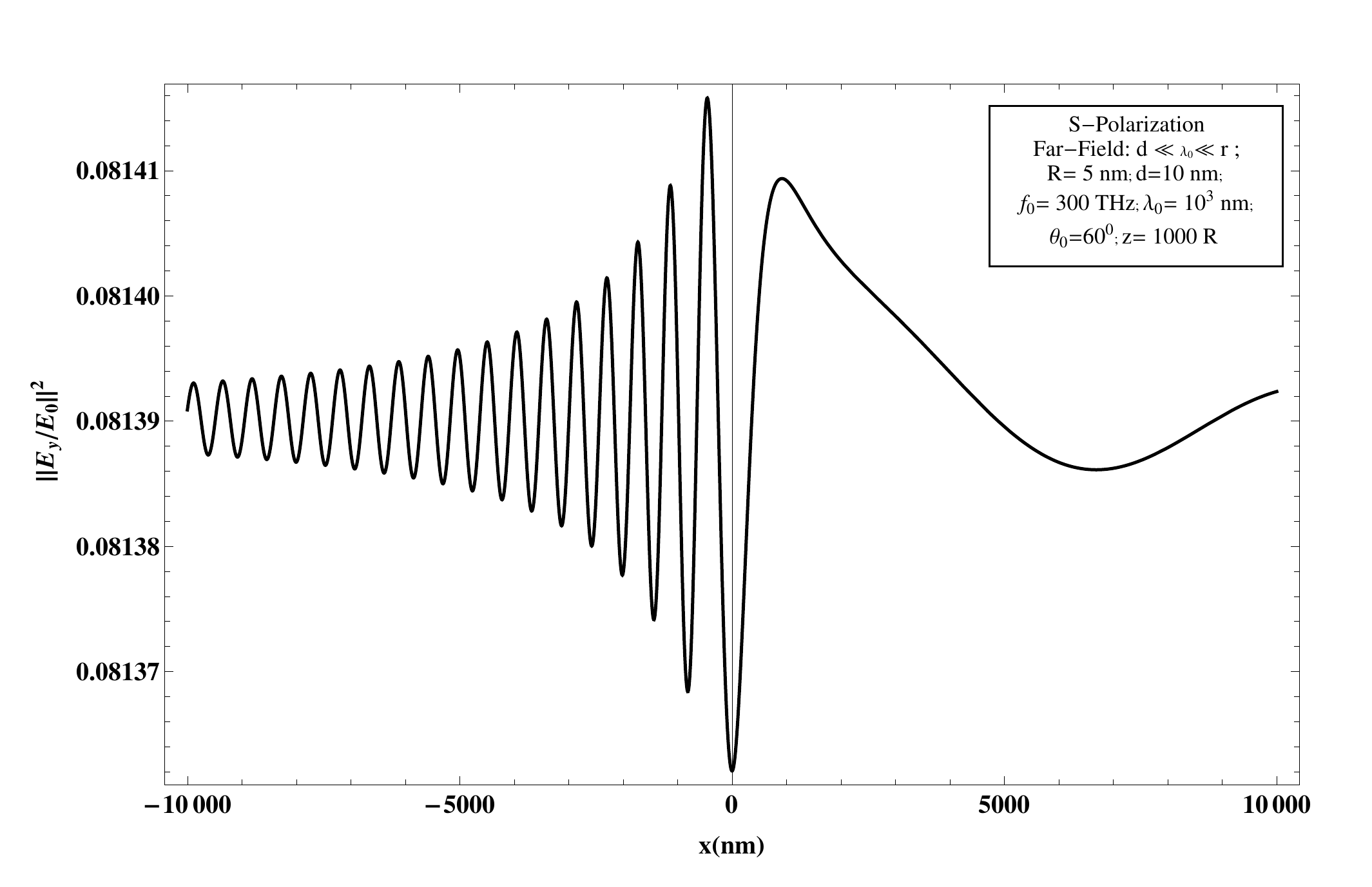}\\
(c)\\
\includegraphics[width=9cm,height=6cm]{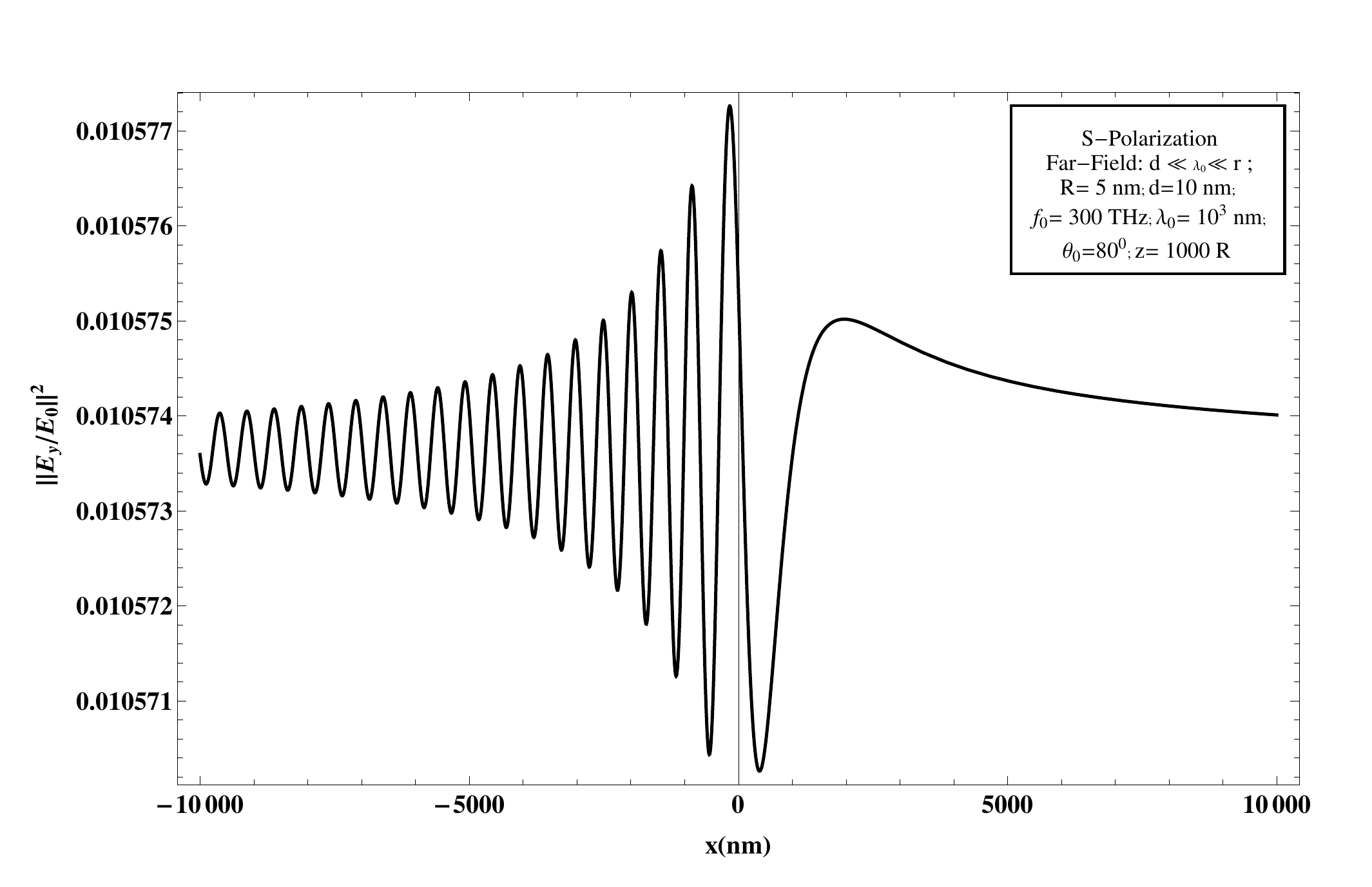}
\caption{$ s $ - polarization - Far-Field, $ z = 1000\,R $; $\mid {E}_{y}(x,y,z;t)/{E}_{0}\mid^{2}$:
(a) $ \theta_{0}=30^{\circ}$ , (b) $ \theta_{0}=60^{\circ}$ and (c) $ \theta_{0}=80^{\circ}$ produced by a perforated 2D plasmonic layer of GaAs as a function of lateral distance $ r_{_{\parallel}}= x\,(y=0)$ from the aperture.}
\label{FIGFFT306080Ey}
\qquad
\end{figure}

The results above are employed here to examine the spatial dependence of the scattered/transmitted wave arising
from an incoming electromagnetic wave at an arbitrary angle of incidence, $\theta_{0}$, in the $ x-z $ plane ($ k_{0y}\equiv 0 $). Detailed computations are carried out for
incident angles of $ 30^{\circ}$, $ 60^{\circ}$ and $ 80^{\circ}$.  The figures below present results for
$\mid E_{x}(x,y,z;t)/E_{0}\mid^{2}$, $\mid E_{y}(x,y,z;t)/E_{0}\mid^{2}$ and $\mid E_{z}(x,y,z;t)/E_{0}\mid^{2}$ in the near, intermediate and
far field diffraction zones and are shown in both 3D and density plots for both $ p-$ and $ s-$polarizations of the incident wave.
In all computations represented in these figures,
we employ the following parameters: $ R =5\,nm $, $ d = 10\,nm $ and $ {f}_{0}= 300\,THz $; the screen is taken to be $ GaAs $ with effective mass $ m^{\ast}=0.067\,m_{0} $ ($ m_{0} $ is the free-electron mass) and density $ n_{3D}=4 \times 10^{21}/cm^{3}$ ($ \varepsilon_{b}^{(3D)}=1 $ is the dielectric constant of the host medium).  For the near, intermediate and far field zones, we fix the $z$-coordinates at $ z=50\,R $, $ z=300\,R $ and $ z=1000\,R $, respectively.  In Figures \ref{FIGNFT30ExEz}-\ref{FIGFFT306080Ey}, we set $y\equiv0$ and $ r_{\parallel}=x$ varies over the indicated zone range, Figures \ref{FIG3DNFR5T30Exz}a-\ref{FIG3DMFR5T306080Ey}a exhibit $3D$ plots of the transmitted power distributions, while Figures \ref{FIG3DNFR5T30Exz}b-\ref{FIG3DMFR5T306080Ey}b provide the associated power density plots, all as functions of both $ x$ and $ y $ for the fixed $z$-values indicated above.

\newpage
\numberwithin{subsection}{section}
\numberwithin{figure}{section}
\textbf{$ p $ - polarization: Near-Field, $ z = 50\,R $; $ \theta_{0}=30^{\circ}$ }
\newpage
\textbf{$ p $ - polarization: Near-Field, $ z = 50\,R $; $ \theta_{0}=60^{\circ}$}
\newpage
\textbf{$ p $ - polarization: Near-Field, $ z = 50\,R $; $ \theta_{0}=80^{\circ}$}
\newpage
\textbf{$ s $ - polarization: Near-Field, $ z = 50\,R $ }
\newpage
\textbf{$ p $ - polarization: Middle-Field, $ z = 300\,R $; $ \theta_{0}=30^{\circ}$ }
\newpage
\textbf{$ p $ - polarization: Middle-Field, $ z = 300\,R $; $ \theta_{0}=60^{\circ}$}
\newpage
\textbf{$ p $ - polarization: Middle-Field, $ z = 300\,R $; $\theta_{0}=80^{\circ}$}
\newpage
\textbf{$ s $ - polarization: Middle-Field, $ z = 300\,R $ }
\newpage
\textbf{$ p $ - polarization: Far-Field, $ z = 1000\,R $; $ \theta_{0}=30^{\circ}$ }
\newpage
\textbf{$ p $ - polarization: Far-Field, $ z = 1000\,R $; $\theta_{0}=60^{\circ}$}
\newpage
\textbf{$ p $ - polarization: Far-Field, $ z = 1000\,R $; $ \theta_{0}=80^{\circ}$}
\newpage
\textbf{$ s $ - polarization: Far-Field, $ z = 1000\,R $ }

\newpage
\numberwithin{figure}{section}
\textbf{$ p $ - polarization: Near-Field, $ z = 50\,R $; $ \theta_{0}=30^{\circ}$ }
\begin{figure}[h]
\centering
\caption{$ p $ - polarization: Near-Field, $ z = 50\,R $; $ \theta_{0}=30^{\circ}$ - Field distribution of GaAs layer in terms of 3D $(a's)$ and density $(b's)$ plots:
$\mid E_{x}(x,y,z;t)/E_{0}\mid^{2}$ and $ \mid E_{z}(x,y,z;t)/E_{0}\mid^{2}$ as functions of $ x $ and $ y $ for fixed $ z $.}
\label{FIG3DNFR5T30Exz}
\end{figure}
\newpage
\textbf{$ p $ - polarization: Near-Field, $ z = 50\,R $; $ \theta_{0}=60^{\circ}$}
\begin{figure}[h]
\centering
\caption{$ p $ - polarization: Near-Field, $ z = 50\,R $; $ \theta_{0}=60^{\circ}$ - Field distribution of GaAs layer in terms of 3D $(a's)$ and density $(b's)$ plots:
$\mid E_{x}(x,y,z;t)/E_{0}\mid^{2}$ and $ \mid E_{z}(x,y,z;t)/E_{0}\mid^{2}$ as functions of $ x $ and $ y $ for fixed $ z $.}
\label{FIG3DNFR5T60Exz}
\end{figure}
\newpage
\textbf{$ p $ - polarization: Near-Field, $ z = 50\,R $; $ \theta_{0}=80^{\circ}$}
\begin{figure}[h]
\centering
\caption{$ p $ - polarization: Near-Field, $ z = 50\,R $; $ \theta_{0}=80^{\circ}$ - Field distribution of GaAs layer in terms of 3D $(a's)$ and density $(b's)$ plots:
$\mid E_{x}(x,y,z;t)/E_{0}\mid^{2}$ and $ \mid E_{z}(x,y,z;t)/E_{0}\mid^{2}$ as functions of $ x $ and $ y $ for fixed $ z $.}
\label{FIG3DNFR5T80Exz}
\end{figure}
\newpage
\textbf{$ s $ - polarization: Near-Field, $ z = 50\,R $}
\begin{figure}[h]
\centering
\caption{{$ s $ - polarization: Near-Field, $ z = 50\,R $ } - Field distribution of GaAs layer in terms of 3D $(a's)$ and density $(b's)$ plots:
$\mid E_{y}(x,y,z;t)/E_{0}\mid^{2}$ (a) $ \theta_{0}=30^{\circ}, 60^{\circ}$ and (b) $ \theta_{0}=30^{\circ}, 60^{\circ}$ as functions of $ x $ and $ y $ for fixed $ z $.}
\label{FIG3DNFR5T306080Ey}
\qquad
\end{figure}
\newpage
\textbf{$ s $ - polarization: Near-Field, $ z = 50\,R $}
\begin{figure}[h]
\centering
\caption{{$ s $ - polarization: Near-Field, $ z = 50\,R $ } - Field distribution of GaAs layer in terms of 3D $(a)$ and density $(b)$ plots:
$\mid E_{y}(x,y,z;t)/E_{0}\mid^{2}$ (a) $ \theta_{0}= 80^{\circ}$ and (b) $ \theta_{0}= 80^{\circ}$ as functions of $ x $ and $ y $ for fixed $ z $.}
\label{FIG3DNFR5T306080Ey1}
\end{figure}
\newpage
\textbf{$ p $ - polarization: Middle-Field, $ z = 300\,R $; $ \theta_{0}=30^{\circ}$}
\begin{figure}[h]
\centering
\caption{$ p $ - polarization: Middle-Field, $ z = 300\,R $; $ \theta_{0}=30^{\circ}$ - Field distribution of GaAs layer in terms of 3D $(a's)$ and density $(b's)$ plots:
$\mid E_{x}(x,y,z;t)/E_{0}\mid^{2}$ and $\mid E_{z}(x,y,z;t)/E_{0}\mid^{2}$ as functions of $ x $ and $ y $ for fixed $ z $.}
\label{FIG3DMFR5T30Exz}
\end{figure}
\newpage
\textbf{$ p $ - polarization: Middle-Field, $ z = 300\,R $; $ \theta_{0}=60^{\circ}$}
\begin{figure}[h]
\centering
\caption{$ p $ - polarization: Middle-Field, $ z = 300\,R $; $ \theta_{0}=60^{\circ}$ - Field distribution of GaAs layer in terms of 3D $(a's)$ and density $(b's)$ plots:
$\mid E_{x}(x,y,z;t)/E_{0}\mid^{2}$ and $ \mid E_{z}(x,y,z;t)/E_{0}\mid^{2}$ as functions of $ x $ and $ y $ for fixed $ z $.}
\label{FIG3DMFR5T60Exz}
\end{figure}
\newpage
\textbf{$ p $ - polarization: Middle-Field, $ z = 300\,R $; $ \theta_{0}=80^{\circ}$}
\begin{figure}[h]
\centering
\caption{$ p $ - polarization: Middle-Field, $ z = 300\,R $; $ \theta_{0}=80^{\circ}$ -Field distribution of GaAs layer in terms of 3D $(a's)$ and density $(b's)$ plots:
$\mid E_{x}(x,y,z;t)/E_{0}\mid^{2}$ and $ \mid E_{z}(x,y,z;t)/E_{0}\mid^{2}$ as functions of $ x $ and $ y $ for fixed $ z $.}
\label{FIG3DMFR5T80Exz}
\end{figure}
\newpage
\textbf{$ s $ - polarization: Middle-Field, $ z = 300\,R $}
\begin{figure}[h]
\centering
\caption{$ s $ - polarization: Middle-Field, $ z = 300\,R $ - Field distribution of GaAs layer in terms of 3D $(a's)$ and density $(b's)$ plots:
$\mid E_{y}(x,y,z;t)/E_{0}\mid^{2}$ (a) $ \theta_{0}=30^{\circ}, 60^{\circ}$ and (b) $ \theta_{0}=30^{\circ}, 60^{\circ}$ as functions of $ x $ and $ y $ for fixed $ z $.}
\label{FIG3DMFR5T306080Ey}
\end{figure}
\newpage
\textbf{$ s $ - polarization: Middle-Field, $ z = 300\,R $}
\begin{figure}[h]
\centering
\caption{$ s $ - polarization: Middle-Field, $ z = 300\,R $ - Field distribution of GaAs layer in terms of 3D $(a)$ and density $(b)$ plots:
$\mid E_{y}(x,y,z;t)/E_{0}\mid^{2}$ (a) $ \theta_{0}=80^{\circ}$ and (b) $ \theta_{0}=80^{\circ}$
as functions of $ x $ and $ y $ for fixed $ z $.}
\label{FIG3DMFR5T306080Ey1}
\end{figure}
\newpage
\textbf{$ p $ - polarization: Far-Field, $ z = 1000\,R $; $ \theta_{0}=30^{\circ}$}
\begin{figure}[h]
\centering
\caption{$ p $ - polarization: Far-Field, $ z = 1000\,R $; $ \theta_{0}=30^{\circ}$ - Field distribution of GaAs layer in terms of 3D $(a's)$ and density $(b's)$ plots:
$\mid E_{x}(x,y,z;t)/E_{0}\mid^{2}$ and $\mid E_{z}(x,y,z;t)/E_{0}\mid^{2}$ as functions of $ x $ and $ y $ for fixed $ z $.}
\label{FIG3DFFR5T30Exz}
\end{figure}
\newpage
\textbf{$ p $ - polarization: Far-Field, $ z = 1000\,R $; $ \theta_{0}=60^{\circ}$}
\begin{figure}[h]
\centering
\caption{$ p $ - polarization: Far-Field, $ z = 1000\,R $; $ \theta_{0}=60^{\circ}$ - Field distribution of GaAs layer in terms of 3D $(a's)$ and density $(b's)$ plots:
$\mid E_{x}(x,y,z;t)/E_{0}\mid^{2}$ and $ \mid E_{z}(x,y,z;t)/E_{0}\mid^{2}$ as functions of $ x $ and $ y $ for fixed $ z $.}
\label{FIG3DFFR5T60Exz}
\end{figure}
\newpage
\textbf{$ p $ - polarization: Far-Field, $ z = 1000\,R $; $ \theta_{0}=80^{\circ}$}
\begin{figure}[h]
\centering
\caption{$ p $ - polarization: Far-Field, $ z = 1000\,R $; $ \theta_{0}=80^{\circ}$ - Field distribution of GaAs layer in terms of 3D $(a's)$ and density $(b's)$ plots:
$\mid E_{x}(x,y,z;t)/E_{0}\mid^{2}$ and $ \mid E_{z}(x,y,z;t)/E_{0}\mid^{2}$ as functions of $ x $ and $ y $ for fixed $ z $.}
\label{FIG3DFFR5T80Exz}
\end{figure}
\newpage
\textbf{$ s $ - polarization: Far-Field, $ z = 1000\,R $}
\begin{figure}[h]
\centering
\caption{$ s $ - polarization: Far-Field, $ z = 1000\,R $ - Field distribution of GaAs layer in terms of 3D $(a's)$ and density $(b's)$ plots:
$\mid E_{y}(x,y,z;t)/E_{0}\mid^{2}$ (a) $ \theta_{0}=30^{\circ}, 60^{\circ}$ and (b) $ \theta_{0}=30^{\circ}, 60^{\circ}$
as functions of $ x $ and $ y $ for fixed $ z $.}
\label{FIG3DFFR5T306080Ey}
\end{figure}
\newpage
\textbf{$ s $ - polarization: Far-Field, $ z = 1000\,R $}
\begin{figure}[h]
\centering
\caption{$ s $ - polarization: Far-Field, $ z = 1000\,R $ - Field distribution of GaAs layer in terms of 3D $(a)$ and density $(b)$ plots:
$\mid E_{y}(x,y,z;t)/E_{0}\mid^{2}$ (a) $ \theta_{0}=80^{\circ}$ and (b) $ \theta_{0}=80^{\circ}$ as functions of $ x $ and $ y $ for fixed $ z $.}
\label{FIG3DFFR5T306080Ey1}
\end{figure}
\section{Conclusions}
In this work we have explored the role of \emph{non}-normal angles of incidence on the transmission of an
electromagnetic waves train through a nano-hole in a thin plasmonic semiconductor screen. This study is
based on our previously constructed [1-5] closed-form dyadic electromagnetic Green's function for a thin
plasmonic/excitonic layer adapted to embody a nano-hole.
The resulting closed-form dyadic Green's function encompasses electromagnetic wave
transmission through both the hole as well as through the screen itself.
This analytic approach involving closed-form solutions of associated integral equations
has facilitated the relatively simple numerical computations exhibited above,
and is not in any way restricted to a metallic screen.   Moreover, our formulation,
which is based on the use of an \textit{integral} equation for the dyadic Green's function
involved automatically incorporates the boundary conditions, which would otherwise need to be addressed explicitly.
It also incorporates the role of the two dimensional plasmon of the thin layer, which is smeared by
its lateral wavenumber dependence.

The calculated results shown in the figures of Section III contrast sharply with the corresponding figures
for normal incidence.  Even for the lowest incident angle of $ 30^{\circ} $ considered in the case of $ p $-polarization,
the near field zone $ (z=50\,R)$ results are highly asymmetric in $ x $ (while the corresponding results for normal incidence
are symmetric).  Such strong asymmetry persists at higher angles of incidence considered $(60^{\circ}, 80^{\circ})$, as may be
seen in Figs.\ref{FIGNFT30ExEz}-\ref{FIGNFT306080Ey} for $ p $-polarization and $ s $-polarization in the near field zone.
Corresponding asymmetric transmission results are exhibited for the middle field zone $(z=300\,R)$ in Figs.\ref{FIGMFT30ExEz}-\ref{FIGMFT306080Ey}
for  $ p $-and $ s $-polarizations at incident angles of $ 30^{\circ}, 60^{\circ}, 80^{\circ}$.  Far field zone ($ z=1000\,R $) transmission
results, also asymmetric, are shown in Figs.\ref{FIGFFT30ExEz}-\ref{FIGFFT306080Ey} for  $ p $- and $ s $-polarizations
at angles of incidence $ 30^{\circ}, 60^{\circ}, 80^{\circ}$.  Further supporting $ 3D $ and density plots are given in
Figs.\ref{FIG3DNFR5T30Exz}a-\ref{FIG3DFFR5T306080Ey1}a and \ref{FIG3DNFR5T30Exz}b-\ref{FIG3DFFR5T306080Ey1}b for the various angles of incidence
and polarizations in the spatial zones considered.

All of the figures exhibit interference fringes due to the superposition of the field transmitted through the nano-hole with
the field transmitted  \textit{directly} through the plasmonic sheet.  At large distances from the nano-hole, the transmission
directly through the plasmonic sheet dominates, and the interference fringes flatten to a uniform level of transmission through
the sheet alone, with the nano-hole contribution negligible.

Finally, it should be noted that the figures show that as the incident angle increases, the axis of the relatively large central transmission
maximum follows it, accompanied by a spatial compression of interference fringe maxima forward of the large central transmission maximum and a spatial
thinning of the fringe maxima behind it.  Moreover, while there is strong asymmetry of electromagnetic transmission with respect to
the $x$-axis (of the $x-z$ plane of incidence), it should be borne in mind that the transmission is fully symmetric with respect to
the $y$-axis (normal to the plane of incidence).  Furthermore, the $ p $-polarization transmission results show strong increase as
incident angle $ \theta_{0}$ increases, mainly in the $ E_{z} $ component.  Although the corresponding $ E_{x}$ component results
decrease as $ \theta_{0}$ increases, the overall combined transmission increases as a function of $ \theta_{0}$.
The $ p $-polarization results for the resultant power transmission, described in terms of \newline
$\mid E_{xz}(x,y,z;t)\mid^{2} \equiv \mid E_{x}(x,y,z;t)\mid^{2} + \mid E_{z}(x,y,z;t)\mid^{2}$ are exhibited in
Figs.\ref{FIG3DNFR5T30Exz}-\ref{FIG3DNFR5T80Exz}, for the near field region and in Figs.\ref{FIG3DMFR5T30Exz}-\ref{FIG3DMFR5T80Exz} for the middle field, also in Figs.\ref{FIG3DFFR5T30Exz}-\ref{FIG3DFFR5T80Exz} for the far field, in both $ 3D $ and density plots in all
cases.   We also find that in the case of $ s $-polarization, the net transmission decreases as $ \theta_{0}$ increases.
All of these results, for both $ p $- and $ s $-polarizations, are consistent with those of Petersson and Smith [20] notwithstanding
the introduction of the $2D$ plasmon of the semiconductor layer accounted for in the present work.
\newpage
\onecolumn
\textbf{$ P $ - polarization: Near-Field, $ z = 50\,R $}
\begin{figure}[h]
\centering
\caption{{$ p $ - polarization: Near-Field, $ z = 50\,R $ } - Transmitted field distribution of GaAs layer for $ \theta_{0}=30^{\circ}$ in terms of 3D $(a)$ and density $(b)$ plots:
$\mid E_{xz}(x,y,z;t)/E_{0}\mid^{2}$ plotted as a function of $ x $ and $ y $ for fixed $ z $.}
\label{FIG3DNFR5T30Exz}
\qquad
\end{figure}
\begin{figure}[h]
\centering
\caption{{$ p $ - polarization: Near-Field, $ z = 50\,R $ } - Transmitted field distribution of GaAs layer for $ \theta_{0}=60^{\circ}$ in terms of 3D $(a)$ and density $(b)$ plots:
$\mid E_{xz}(x,y,z;t)/E_{0}\mid^{2}$ plotted as a function of $ x $ and $ y $ for fixed $ z $.}
\label{FIG3DNFR5T60Exz}
\qquad
\end{figure}
\begin{figure}[h]
\centering
\caption{{$ p $ - polarization: Near-Field, $ z = 50\,R $ } - Transmitted field distribution of GaAs layer for $ \theta_{0}=80^{\circ}$ in terms of 3D $(a)$ and density $(b)$ plots:
$\mid E_{xz}(x,y,z;t)/E_{0}\mid^{2}$ plotted as a function of $ x $ and $ y $ for fixed $ z $.}
\label{FIG3DNFR5T80Exz}
\end{figure}
\newpage
\textbf{$ P $ - polarization: Middle-Field, $ z = 300\,R $}
\begin{figure}[h]
\centering
\caption{{$ p $ - polarization: Middle-Field, $ z = 300\,R $ } - Transmitted field distribution of GaAs layer for $ \theta_{0}=30^{\circ}$ in terms of 3D $(a)$ and density $(b)$ plots:
$\mid E_{xz}(x,y,z;t)/E_{0}\mid^{2}$ plotted as a function of $ x $ and $ y $ for fixed $ z $.}
\label{FIG3DMFR5T30Exz}
\qquad
\end{figure}
\begin{figure}[h]
\centering
\caption{{$ p $ - polarization: Middle-Field, $ z = 300\,R $ } - Transmitted field distribution of GaAs layer for $ \theta_{0}=60^{\circ}$ in terms of 3D $(a)$ and density $(b)$ plots:
$\mid E_{xz}(x,y,z;t)/E_{0}\mid^{2}$ plotted as a function of $ x $ and $ y $ for fixed $ z $.}
\label{FIG3DMFR5T60Exz}
\qquad
\end{figure}
\begin{figure}[h]
\centering
\caption{{$ p $ - polarization: Middle-Field, $ z = 300\,R $ } - Transmitted field distribution of GaAs layer for $ \theta_{0}=80^{\circ}$ in terms of 3D $(a)$ and density $(b)$ plots:
$\mid E_{xz}(x,y,z;t)/E_{0}\mid^{2}$ plotted as a function of $ x $ and $ y $ for fixed $ z $.}
\label{FIG3DMFR5T80Exz}
\end{figure}
\newpage
\textbf{$ P $ - polarization: Far-Field, $ z = 1000\,R $}
\begin{figure}[h]
\centering
\caption{{$ p $ - polarization: Far-Field, $ z = 1000\,R $ } - Transmitted field distribution of GaAs layer for $ \theta_{0}=30^{\circ}$ in terms of 3D $(a)$ and density $(b)$ plots:
$\mid E_{xz}(x,y,z;t)/E_{0}\mid^{2}$ plotted as a function of $ x $ and $ y $ for fixed $ z $.}
\label{FIG3DFFR5T30Exz}
\qquad
\end{figure}
\begin{figure}[h]
\centering
\caption{{$ p $ - polarization: Far-Field, $ z = 1000\,R $ } - Transmitted field distribution of GaAs layer for $ \theta_{0}=60^{\circ}$ in terms of 3D $(a)$ and density $(b)$ plots:
$\mid E_{xz}(x,y,z;t)/E_{0}\mid^{2}$ plotted as a function of $ x $ and $ y $ for fixed $ z $.}
\label{FIG3DFFR5T60Exz}
\qquad
\end{figure}
\begin{figure}[h]
\centering
\caption{{$ p $ - polarization: Far-Field, $ z = 1000\,R $ } - Transmitted field distribution of GaAs layer for $ \theta_{0}=80^{\circ}$ in terms of 3D $(a)$ and density $(b)$ plots:
$\mid E_{xz}(x,y,z;t)/E_{0}\mid^{2}$ plotted as a function of $ x $ and $ y $ for fixed $ z $.}
\label{FIG3DFFR5T80Exz}
\end{figure}
\twocolumn
\appendices
\section{Matrix Elements of $\widehat{\Omega}^{-1}$ and $\widehat{G}_{3D}(\vec{k}_{\parallel},z,0;\omega)$}\label{App:AppendixA}
The elements of $\widehat{\Omega}^{-1}$ are given by(notation: $ a_{0}^{2}=k_{z}^{2}-
\frac{2 i k_{z}}{d}$)
\small
\begin{eqnarray}\label{A.1}
  [\widehat{\Omega}^{-1}]_{_{11}} =  \frac{\left[1+ \left(\frac{\gamma}{2\,i\,k_{z}}
  \right) \left( 1- \frac{k_{y}^{2}}{q_{\omega}^{2}}\right)
 \right]}{\left[\left(1+\left(\frac{\gamma}{2\,i\,k_{z}}\right)\right)\left(1+ \left(\frac{\gamma}{2\,i\,k_{z}}\right) \left( 1- \frac{k_{\parallel}^{2}}{q_{\omega}^{2}}\right) \right)
 \right]},
\end{eqnarray}
\begin{eqnarray}\label{A.2}
  [\widehat{\Omega}^{-1}]_{_{12}} =  \frac{\left[\left(\frac{\gamma}{2\,i\,k_{z}}\right) \left(\frac{k_{x}\,k_{y}}{q_{\omega}^{2}} \right)
 \right]}{\left[\left(1+\left(\frac{\gamma}{2\,i\,k_{z}}\right)\right)\left(1+ \left(\frac{\gamma}{2\,i\,k_{z}}\right) \left( 1- \frac{k_{\parallel}^{2}}{q_{\omega}^{2}}\right) \right)
 \right]},
\end{eqnarray}
\begin{equation}\label{A.3}
    [\widehat{\Omega}^{-1}]_{_{21}} = [\widehat{\Omega}^{-1}]_{_{12}},
 \end{equation}
\begin{equation}\label{A.4}
    [\widehat{\Omega}^{-1}]_{_{31}} = [\widehat{\Omega}^{-1}]_{_{13}}=[\widehat{\Omega}^{-1}]_{_{32}}=[\widehat{\Omega}^{-1}]_{_{23}}=0,
\end{equation}
\begin{eqnarray}\label{A.5}
 [\widehat{\Omega}^{-1}]_{_{22}} = \frac{\left[1+ \left(\frac{\gamma}{2\,i\,k_{z}}\right) \left( 1- \frac{k_{x}^{2}}{q_{\omega}^{2}}\right)
 \right]}{\left[\left(1+\left(\frac{\gamma}{2\,i\,k_{z}}\right)\right)\left(1+ \left(\frac{\gamma}{2\,i\,k_{z}}\right) \left( 1- \frac{k_{\parallel}^{2}}{q_{\omega}^{2}}\right) \right)
 \right]},
\end{eqnarray}
\normalsize
and
\small
\begin{eqnarray}\label{A.6}
  [\widehat{\Omega}^{-1}]_{_{33}} = \frac{1}{\left[1+ \left(\frac{\gamma}{2\,i\,k_{z}}\right) \left( 1- \frac{a_{0}^{2}}{q_{\omega}^{2}}\right)
 \right]}.
\end{eqnarray}
\normalsize
\section{}\label{App:AppendixB}
Furthermore the elements of ${\widehat{G}_{3D}}({\vec{k}}_{\parallel};z,0;
\omega)$ are given by the matrix ${{G}_{3D}}^{\,ij}({\vec{k}}_{
\parallel};z,0;\omega)$
as:
\begin{equation}\label{B.1}
{G}_{3D}^{xx}\left({\vec{k}}_{\parallel};z,0;\omega\right) =\,-\,\frac{e^{i\,k_{z}\mid z\mid }}{2\,i\,k_{z}}\left(1-\frac{k_{x}^{2}}{q_{\omega}^{2}}\right),
\end{equation}
\begin{equation}\label{B.2}
{G}_{3D}^{yy}\left({\vec{k}}_{\parallel};z,0;\omega\right)=\,-\,\frac{e^{i\,k_{z}\mid z\mid }}{2\,i\,k_{z}}\left(1-\frac{k_{y}^{2}}{q_{\omega}^{2}}\right),
\end{equation}
\begin{equation}\label{B.3}
{G}_{3D}^{zz}\left({\vec{k}}_{\parallel};z,0;\omega\right)=\,-\,\frac{e^{i\,k_{z}\mid z\mid }}{2\,i\,k_{z}}\left(1-\frac{k_{z}^{2}- 2 i k_{z} \delta (z)}{q_{\omega}^{2}}\right),
\end{equation}
\begin{equation}\label{B.4}
{G}_{3D}^{xy}\left({\vec{k}}_{\parallel};z,0;\omega\right)=\,+\,\frac{e^{i\,k_{z}\mid z\mid }}{2\,i\,k_{z}}\left(\frac{k_{x}\,k_{y}}{q_{\omega}^{2}}\right),
\end{equation}
\begin{equation}\label{B.5}
{G}_{3D}^{xz}\left({\vec{k}}_{\parallel};z,0;\omega\right)=\,+\,\frac{e^{i\,k_{z}\mid z\mid }}{2\,i\,k_{z}}\left(\frac{k_{x}\,k_{z}sgn(z)}{q_{\omega}^{2}}\right),
\end{equation}
\begin{equation}\label{B.6}
{G}_{3D}^{yz}\left({\vec{k}}_{\parallel};z,0;\omega\right)=\,+\,\frac{e^{i\,k_{z}\mid z\mid }}{2\,i\,k_{z}}\left(\frac{k_{y}\,k_{z}sgn(z)}{q_{\omega}^{2}}\right),
\end{equation}
\begin{equation}\label{B.7}
{G}_{3D}^{yx}\left({\vec{k}}_{\parallel};z,0;\omega\right) = {G}_{3D}^{xy}\left({\vec{k}}_{\parallel};z,0;\omega\right),
\end{equation}
\begin{equation}\label{B.8}
{G}_{3D}^{xz}\left({\vec{k}}_{\parallel};z,0;\omega\right) = {G}_{3D}^{zx}\left({\vec{k}}_{\parallel};z,0;\omega\right),
\end{equation}
\begin{equation}\label{B.9}
{G}_{3D}^{yz}\left({\vec{k}}_{\parallel};z,0;\omega\right) = {G}_{3D}^{zy}\left({\vec{k}}_{\parallel};z,0;\omega\right).
\end{equation}

Therefore,  Eq.(\ref{A2.16}) may rewritten as
\begin{equation}\label{B.10}
   {\widehat{G}_{fs}}({\vec{k}}_{\parallel};z,0;\omega)={\widehat{G}_{3D}}({
   \vec{k}}_{\parallel};z,0;\omega)\,{\widehat{\Omega}}^{-1}.
\end{equation}

\section{Matrix Elements of $\widehat{G}_{fs}$ $ \longmapsto $ ${G}_{fs}^{ij}(\vec{k}_{\parallel},z,0;\omega)$}\label{App:AppendixC}
\tiny
\begin{eqnarray}\label{C.1}
{G}_{fs}^{xx}\left({\vec{k}}_{\parallel};z,0;\omega\right)&=&\,-\,\frac{e^{i\,k_{z}\mid z\mid }}{2\,i\,k_{z}}
\nonumber\\
&\times&
\left[\frac{1}{D_{1}}\left\{\left(1-\frac{k_{x}^{2}}{q_{\omega}^{2}}\right) +\left(\frac{\gamma}{2\,i\,k_{z}}\right)
\left(1-\frac{k_{\parallel}^{2}}{q_{\omega}^{2}}\right)\right\}\right],
\end{eqnarray}
\begin{eqnarray}\label{C.2}
{G}_{fs}^{yy}\left({\vec{k}}_{\parallel};z,0;\omega\right)&=&\,-\,\frac{e^{i\,k_{z}\mid z\mid }}{2\,i\,k_{z}}
\nonumber\\
&\times&
\left[\frac{1}{D_{1}}\left\{\left(1-\frac{k_{y}^{2}}{q_{\omega}^{2}}\right) +\left(\frac{\gamma}{2\,i\,k_{z}}\right)
\left(1-\frac{k_{\parallel}^{2}}{q_{\omega}^{2}}\right)\right\}\right],
\end{eqnarray}
\normalsize
\small
\begin{equation}\label{C.3}
{G}_{fs}^{zz}\left({\vec{k}}_{\parallel};z,0;\omega\right)=\,-\,\frac{e^{i\,k_{z}\mid z\mid }}{2\,i\,k_{z}}\left[\frac{1}{D_{2}}\left\{\left(1-\frac{k_{z}^{2}- 2 i k_{z} \delta (z)}{q_{\omega}^{2}}\right)\right\}\right],
\end{equation}
\begin{equation}\label{C.4}
{G}_{fs}^{xy}\left({\vec{k}}_{\parallel};z,0;\omega\right)=\,+\,\frac{e^{i\,k_{z}\mid z\mid }}{2\,i\,k_{z}}\left[\frac{1}{D_{1}}\left(\frac{k_{x}\,k_{y}}{q_{\omega}^{2}}\right)\right],
\end{equation}
\begin{equation}\label{C.5}
{G}_{fs}^{xz}\left({\vec{k}}_{\parallel};z,0;\omega\right)
=\,+\,\frac{e^{i\,k_{z}\mid z\mid }}{2\,i\,k_{z}}\left[\frac{1}{D_{2}}\left(\frac{k_{x}\,k_{z}sgn(z)}{q_{\omega}^{2}}\right)\right],
\end{equation}
\begin{equation}\label{C.6}
{G}_{fs}^{yz}\left({\vec{k}}_{\parallel};z,0;\omega\right)=\,+\,\frac{e^{i\,k_{z}\mid z\mid }}{2\,i\,k_{z}}\left[\frac{1}{D_{2}}\left(\frac{k_{y}\,k_{z}sgn(z)}{q_{\omega}^{2}}\right)\right],
\end{equation}
\begin{equation}\label{C.7}
{G}_{fs}^{yx}\left({\vec{k}}_{\parallel};z,0;\omega\right) = {G}_{fs}^{xy}\left({\vec{k}}_{\parallel};z,0;\omega\right),
\end{equation}
\begin{eqnarray}\label{C.8}
{G}_{fs}^{zx}\left({\vec{k}}_{\parallel};z,0;\omega\right)&=&\frac{e^{i\,k_{z}\mid z\mid }}{2\,i\,k_{z}}
\nonumber\\
&\times&
\left[\frac{1}{D_{1}}\left(\left(1+\left(\frac{\gamma}{2
\,i\,k_{z}}\right)\right)\frac{k_{z}\,k_{x}sgn(z)}{q_{\omega}^{2}}\right)\right],
\nonumber\\
\end{eqnarray}
\begin{eqnarray}\label{C.9}
{G}_{fs}^{zy}\left({\vec{k}}_{\parallel};z,0;\omega\right)
&=&\frac{e^{i\,k_{z}\mid z\mid }}{2\,i\,k_{z}}
\nonumber\\
&\times&
\left[\frac{1}{D_{1}}\left(\left(1+\left(\frac{
\gamma}{2\,i\,k_{z}}\right)\right)\frac{k_{z}\,k_{y}sgn(z)}{q_{\omega}^{2}}\right)\right],
\nonumber\\
\end{eqnarray}
\begin{equation}\label{C.10}
{G}_{fs}^{xz}\left({\vec{k}}_{\parallel};z,0;\omega\right) = {G}_{fs}^{zx}\left({\vec{k}}_{\parallel};z,0;\omega\right),
\end{equation}
and
\begin{equation}\label{C.11}
   D_{1}=\left(1+\left(\frac{\gamma}{2\,i\,k_{z}}\right)\right)\left[1+ \left(\frac{\gamma}{2\,i\,k_{z}}\right) \left( 1- \frac{k_{\parallel}^{2}}{q_{\omega}^{2}}\right) \right],
\end{equation}
\begin{equation}\label{C.12}
   D_{2}=\left[1+ \left(\frac{\gamma}{2\,i\,k_{z}}\right) \left( 1-
   \frac{a_{0}^{2}}{q_{\omega}^{2}}\right) \right].
\end{equation}
\normalsize
It should be noted that, in the text above, we choose the coordinate system such that
$ k_{y}\equiv 0 $.  In this case, $ k_{\parallel}=k_{x} $ and $ k_{z}^{2}=q_{\omega}^{2}-
k_{x}^{2}$.  Moreover, with $ k_{y}=0 $, $
{\widehat{G}_{3D}}({\vec{k}}_{\parallel};0,0;\omega)$ becomes diagonal.

\section{Matrix Elements of $\widehat{T}^{0}$ }\label{App:AppendixD}
Since $ \widehat{\overline{\overline{G}}}_{fs}(\vec{k}_{0_{\parallel}};0,0;
\omega_{0})$ is diagonal, we have
\begin{eqnarray}\label{D.1}
  {\widehat{T}^{0}}  & = &
  \begin{bmatrix}
  {G}^{xx} & 0 & {G}^{xz} \\
  0 & {G}^{yy} & 0 \\
  {G}^{zx} & 0 & {G}^{zz}
 \end{bmatrix}.
\nonumber\\
&\times&
\begin{bmatrix}
  1+\gamma_{0}{\overline{\overline{G}}}_{fs}^{\,xx} & 0 & 0 \\
  0 & 1+\gamma_{0}{\overline{\overline{G}}}_{fs}^{\,yy} & 0 \\
  0 & 0 & 1+\gamma_{0}{\overline{\overline{G}}}_{fs}^{\,zz}
 \end{bmatrix}
\end{eqnarray}
or
\begin{eqnarray}\label{D.2}
  {\widehat{T}^{0}}  & = &
\begin{bmatrix}
  {T}_{xx}^{0} & 0 & {T}_{xz}^{0} \\
  0 & {T}_{yy}^{0} & 0 \\
  {T}_{zx}^{0} & 0 & {T}_{zz}^{0}
 \end{bmatrix}
\end{eqnarray}
where the matrix elements are given by
\begin{equation}\label{D.3}
\left\{
         \begin{array}{ll}
            {T}_{xx}^{0} = & \hbox{$ G^{xx}\left[ 1\,+\,\gamma_{0}\,\overline{
            \overline{G}}_{fs}^{\,xx}\right]$ ; (a)} \\
             {T}_{zx}^{0} = & \hbox{$ G^{zx}\left[ 1\,+\,\gamma_{0}\,\overline{
             \overline{G}}_{fs}^{\,xx}\right]$ ; (b)} \\
            {T}_{yy}^{0} = & \hbox{$ G^{yy}\left[ 1\,+\,\gamma_{0}\,\overline{
            \overline{G}}_{fs}^{\,yy}\right]$ ; (c)} \\
             {T}_{xz}^{0} = & \hbox{$ G^{xz}\left[ 1\,+\,\gamma_{0}\,\overline{
             \overline{G}}_{fs}^{\,zz}\right]$ ; (d)} \\
            {T}_{zz}^{0} = & \hbox{$ G^{zz}\left[ 1\,+\,\gamma_{0}\,\overline{
            \overline{G}}_{fs}^{\,zz}\right]$ . (e)}
         \end{array}
       \right.
\end{equation}
In order to facilitate the calculations of the above matrix elements, it is important to note that
$ \widehat{G}=\widehat{G}(\vec{r}_{\parallel},0;z,0;\omega_{0}) $
and $ \widehat{\overline{\overline{G}}}_{fs}= \widehat{\overline{\overline{G}}}_{fs}(\vec{k}_{0_{\parallel}};0,0;\omega_{0})$ are
 evaluated at the incident wave frequency $
\omega_{0}$.  Therefore, the matrix elements of the dyad $ {\widehat{T}^{0}} $
are given by (note that $\theta_{0}$ is the angle of incident)
\begin{equation}\label{D.4}
    {T}_{xx}^{0} = G_{fs}^{xx}\left[\frac{1\,+\,\gamma_{0}\,\overline{
    \overline{G}}_{fs}^{\,xx}}{1\,
    +\,\beta_{0}\,{G}_{fs0}^{\,xx}}\right] \,\,\,\,
\end{equation}
where
\begin{equation}\label{D.4I}
    \left\{
       \begin{array}{ll}
         \overline{\overline{G}}_{fs}^{\,xx} \equiv\, & \hbox{$\frac{\,-\,\cos(\theta_{0})}{2\,i\,\left[q_{\omega_{0}}\,+\,\Gamma_{0}\,\cos(\theta_{0})\right]}$;    (a)} \\
         {G}_{fs0}^{\,xx} \equiv\, & \hbox{$ {G}_{fs0}^{\,xx}(0-0;0,0;\omega_{0}) $;     (b)} \\
         {G}_{fs}^{\,xx} \equiv\, & \hbox{$ {G}_{fs}^{\,xx}(\vec{r}_{\parallel},0;z,0;\omega_{0})$.    (c)}
       \end{array}
     \right.
\end{equation}
\begin{equation}\label{D.5}
    {T}_{yy}^{0} = G_{fs}^{yy}\left[\frac{1\,+\,\gamma_{0}\,\overline{\overline{G}}_{fs}^{\,yy}}{1\,+\,\beta_{0}\,{G}_{fs0}^{\,yy}}\right] \,\,\,\,
\end{equation}
where
\begin{equation}\label{D.5I}
    \left\{
       \begin{array}{ll}
         \overline{\overline{G}}_{fs}^{\,yy} \equiv\, & \hbox{$\frac{\,-\,1}{2\,i\,\left[\,\Gamma_{0}\,+\,q_{\omega_{0}}\,\cos(\theta_{0})
         \right]}$;    (a)} \\
         {G}_{fs0}^{\,yy} \equiv\, & \hbox{$ {G}_{fs0}^{\,yy}(0-0;0,0;\omega_{0}) $;     (b)} \\
         {G}_{fs}^{\,yy} \equiv\, & \hbox{$ {G}_{fs}^{\,yy}(
         \vec{r}_{\parallel},0;z,0;\omega_{0})$.    (c)}
       \end{array}
     \right.
\end{equation}
\begin{equation}\label{D.6}
    {T}_{zz}^{0} = G_{fs}^{zz}\left[\frac{1\,+\,\gamma_{0}\,\overline{\overline{G}}_{fs}^{\,zz}}{1\,+\,\beta_{0}\,{G}_{fs0}^{\,zz}}\right] \,\,\,\,
\end{equation}
where
\small
\begin{equation}\label{D.6I}
    \left\{
       \begin{array}{ll}
         \overline{\overline{G}}_{fs}^{\,zz} \equiv\, & \hbox{$  \frac{\,-\,1}{2\,i} \left\{\frac{q_{\omega_{0}}\,\sin^{2}(\theta_{0})+2\,i\,\delta(0)\cos(\theta_{0})}{\cos(\theta_{0})\,
\left[\,q_{\omega_{0}}^{2}\,+\,2\,i\,\delta(0)\Gamma_{0}\right]\,+\,\Gamma_{0}\,q_{\omega_{0}}\,\sin^{2}(\theta_{0})}\right\}$;    (a)} \\
         {G}_{fs0}^{\,zz} \equiv\, & \hbox{$ {G}_{fs0}^{\,zz}(0-0;0,0;\omega_{0}) $;     (b)} \\
         {G}_{fs}^{\,zz}\equiv\, & \hbox{$ {G}_{fs}^{\,zz}(\vec{r}_{\parallel},0;z,0;\omega_{0})$.    (c)}
       \end{array}
     \right.
\end{equation}
\normalsize
\begin{equation}\label{D.7}
    {T}_{xz}^{0} = G_{fs}^{xz}\left[\frac{1\,+\,\gamma_{0}\,\overline{\overline{G}}_{fs}^{\,zz}}{1\,+\,\beta_{0}\,{G}_{fs0}^{\,zz}}\right] \,\,\,\,
\end{equation}
where
\small
\begin{equation}\label{D.7I}
    \left\{
       \begin{array}{ll}
         \overline{\overline{G}}_{fs}^{\,zz} \equiv\, & \hbox{$  \frac{\,-\,1}{2\,i} \left\{\frac{q_{\omega_{0}}\,\sin^{2}(\theta_{0})\,+2\,i\,\delta(0)\cos(\theta_{0})}{\cos(\theta_{0})\,
\left[\,q_{\omega_{0}}^{2}\,+\,2\,i\,\delta(0)\,\Gamma_{0}\right]\,+\,\Gamma_{0}\,q_{\omega_{0}}\,\sin^{2}(\theta_{0})}\right\}$;    (a)} \\
         {G}_{fs0}^{\,zz} \equiv\, & \hbox{$ {G}_{fs0}^{\,zz}(0-0;0,0;\omega_{0}) $;     (b)} \\
         {G}_{fs}^{\,xz} \equiv\, & \hbox{$ {G}_{fs}^{\,xz}(\vec{r}_{\parallel},0;z,0;\omega_{0})$.    (c)}
       \end{array}
     \right.
\end{equation}
\normalsize
and
\begin{equation}\label{D.8}
    {T}_{zx}^{0} = G_{fs}^{zx}\left[\frac{1\,+\,\gamma_{0}\,\overline{\overline{G}}_{fs}^{\,xx}}{1\,+\,\beta_{0}\,{G}_{fs0}^{\,xx}}\right] \,\,\,\,
\end{equation}
where
\begin{equation}\label{D.8I}
    \left\{
       \begin{array}{ll}
         \overline{\overline{G}}_{fs}^{\,xx} \equiv\, & \hbox{$\frac{\,-\,\cos(\theta_{0})}{2\,i\,\left[q_{\omega_{0}}\,+\,\Gamma_{0}\,\cos(\theta_{0})\right]}$;    (a)} \\
         {G}_{fs0}^{\,xx} \equiv\, & \hbox{$ {G}_{fs0}^{\,xx}(0-0;0,0;\omega_{0}) $;     (b)} \\
         {G}_{fs}^{\,zx} \equiv\, & \hbox{$ {G}_{fs}^{\,zx}(\vec{r}_{\parallel},0;z,0;\omega_{0})$.    (c)}
       \end{array}
     \right.
\end{equation}

\section*{Acknowledgment}

D. Miessein gratefully acknowledges support by the AGEP program of the NSF; also the assistance of Prof. M. L. Glasser, Dr. Andrii Iurov and Dr. Nan Chen.


\begin{thebibliography}{1}
\bibitem{Dez_2015} N.J. M. Horing, D\'{e}sir\'{e} Miessein and G. Gumbs, J. Opt. Soc. Am. A 32, 1184-1198 (2015).
\bibitem{Dez_thesis} D\'{e}sir\'{e} Miessein, Ph.D Thesis
\bibitem{Horing_2007} N.J. M. Horing, et al., J. Optical Soc. Amer. B. 24, 2428 (2007).
\bibitem{Horing_2015} N.J. M. Horing, " Quantum Statistical Fields Theory: Schwinger's Variational Method", Oxford University Press, in press.
\bibitem{Horing_2008} N.J. M. Horing , IEEE Sensors J. Vol.8, N0.6, 771 (2008).
\bibitem{Bethe_1944} A. Bethe, " Theory of Diffraction by Small Holes ", Phys. Rev. 66, 163,(1944 ).
\bibitem{Levine_1950} Levine and J. Schwinger, Commun. Pure and Appl. Math. 3, 355-391(1950).
\bibitem{Levine_1948} Levine and J. Schwinger, Phys. Rev. 74, 958(1948).
\bibitem{Levine_1949} Levine and J. Schwinger, Phys. Rev. 75, 1423(1949).
\bibitem{Tai_1994} Chen-To Tai, " Dyadic Green's Functions in Electromagnetic Theory", Intext Educational Publishers (1971); reprinted as
            " Dyadic Green's Functions in Electromagnetic Theory", IEEE Press: Piscataway, NJ, (1994).
\bibitem{Tai_1997} Chen-To Tai, " General Vector and Dyadic Analysis: Applied Mathematics in Field Theory",2nd Edition, Wiley-IEEE Press (April 15, 1997).
\bibitem{Collin_1991} Robert E. Collin, " Field Theory of Guided Waves", 2nd Edition, IEEE Press (1991).
\bibitem{Chew_1995} Weng Cho Chew, " Waves and Fields In Inhomogeneous Media" IEEE, Inc, (1995).
\bibitem{Genet_2007} C. Genet and T. Ebbesen, Nature, 445, 390(2007).
\bibitem{Kukh_2006} S.V. Kukhlevsky, M. Mechler, O. Samek, K. Janssens, " Analytical model of the enhanced light transmission through subwavelength metal slits: Green's function formalism versus Rayleigh's expansion ", Appl. Phys. B: Lasers and Optics 84:19-24 (2006).
\bibitem{Kukh_2004} S.V. Kukhlevsky, M. Mechler, L. Csapo, K. Janssens, and O. Samek, Phys. Rev. B 70, 195428 (2004).
\bibitem{Neerhoff_1973} F. L. Neerhoff G. Mur, Appl. Sci. Res. 28, 73 (1973).
\bibitem{tb:integral_1980} Table of Integrals, Series, and Products, I.S. Gradshteyn and I.M. Ryzhik (6th Ed.) Academic Press, New York (1980).
\bibitem{Arnoldus_2002} Henk F. Arnoldus and John T. Foley , " Traveling and evanescent parts of the electromagnetic Green's tensor", J. Optical Soc. Amer. A. 19, 1701 (2002).
\bibitem{Glen_2004} L. E. Richard Petersson and Glenn S. Smith, J. Optical Soc. Amer. A. 21, 975-980 (2004).
\end{thebibliography}
\end{document}